Masterarbeit im Studiengang Physik (M.Sc.)

# Quality-of-Transmission Estimation in Physical Impairment Aware Flexible Optical Networks

vorgelegt von

## Joschua Dilly

Matrikel-Nr. 345235

an der Fakultät II

Mathematik und Naturwissenschaften

der Technischen Universität Berlin

1. Gutachter: Prof. Dr. Martin Schell,
TU Berlin, Fraunhofer Heinrich-Hertz-Institut

2. Gutachter: Dr. Carsten Schmidt-Langhorst,
Fraunhofer Heinrich-Hertz-Institut

Berlin, August 2015

# Eidesstattliche Erklärung

Hiermit erkläre ich, dass ich die vorliegende Arbeit selbstständig und eigenhändig sowie ohne unerlaubte fremde Hilfe und ausschließlich unter Verwendung der aufgeführten Quellen und Hilfsmittel angefertigt habe.

Berlin, den 10 August 2015

______________________________
                Joschua Dilly

# Zusammenfassung

Zur Abschätzung der Übertragungsqualität in flexiblen optischen Netzen mit Leistungskontrolle wird in dieser Arbeit die Entstehung nichtlinearen Rauschens untersucht. Im Besonderen werden die nichtlinearen Störungen, die während der Übertragung durch Nachbarkanäle am untersuchten Kanal verursacht werden, betrachtet. Als ausschlaggebende Parameter werden in diesem Zusammenhang das Modulationsformat der Nachbarkanäle, ihre akkumulierte chromatische Dispersion, die Frequenzabstände der Kanäle und die Länge der Übertragungsabschnitte identifiziert. Aus den Ergebnissen numerischer Simulationen wird das nichtlineare Rauschen unter Einfluss dieser Parameter bestimmt und mit den Resultaten der aktuellen analytischen Modelle „Gaussian Noise Model" und „Enhanced Gaussian Noise Model" verglichen. Zusätzlich zur Rauschleistung liefern die Simulationen auch Aussagen über den Anteil an Phasen- und Zirkularrauschen und die Korrelationszeit des Phasenrauschens. Diese Ergebnisse werden genutzt zur Abschätzung des Einflusses auf die Übertragungsqualität. Eine Interpretation erfolgt außerdem mit Hilfe des „Pulse Collision Pictures".

# Abstract

In this thesis the creation of nonlinear interference noise (NLIN) in the context of impairment aware flexible optical networks is investigated to estimate transmission quality. In particular, the nonlinear interference of neighboring channels (interferer) during transmission on a channel under test is studied. The modulation format of the interferer, the accumulated chromatic dispersion of the interferer, the span length and channel spacing are identified as the parameter influencing the generation of NLIN in the context of flexible optical networks. Estimation of the NLIN is done based on the evaluation of numerical simulation results and compared to recent analytical transmission models, namely the Gaussian noise model and the enhanced Gaussian noise model. In addition to nonlinear noise power, the simulations also yield information about the phase- and circular-noise contributions as well as the correlation time of the phase noise. The results are evaluated in the context of the pulse-collision picture and the influence on transmission quality is discussed.

# Contents

# 1  Introduction

In contrast to static point-to-point networks, where all channels start at the same transmitter, co-propagate along a link and arrive at the same destination, a novel trend in optical communication systems is enhancing the architectures to allow *colorless*, *directionless and contentionless* (CDC, [1]–[4]) switching at the nodes of the link, i.e. being able to route any wavelength over any port to any direction. Modern r*econfigurable optical add/drop multiplexer* (ROADMs) are used for this purpose. This leads to networks in which data packages can originate at multiple locations, travel any path and to any destination, while being located at any part of the spectral range provided by the system, allowing optimal usage of the spectral width as well as the geometries of the system.

A network fulfilling these requirements is called flexible optical network. ***Figure 1.1*** shows a simple example of such a set-up: here, three different channels, blue, red and green, are transmitted along the paths $\overline{AC}$, $\overline{DC}$ and $\overline{BE}$ respectively. Each channel has its individual spectral width and modulation format (depicted by the curves over frequency and the symbolized constellation on top). In course of time, the routes may need to change, i.e. data might be send from A to D, instead of C. As the ROADMs are reconfigurable they allow for such an alteration. The figure also shows, that the set of channels propagating together along sections of the link can change and information about neighboring channels cannot be obtained at neither transmitter (e.g. B) nor receiver (e.g. E) side. In addition to rerouting, the electrical bandwidth of the receiver also limits the number of received channels. Hence, the recipient of a signal in a flexible optical network has only information about the channel destined to his or her location. Applying mitigation schemes, such as backpropagation, will therefore remove linear impairments caused by the link and non-linear perturbations induced by the channel itself, but the influences generated by neighboring channels stay elusive. Hence, one of the most limiting factor on the performance of an optical transmission system is inter-channel *nonlinear interference* (NLI, [5]).

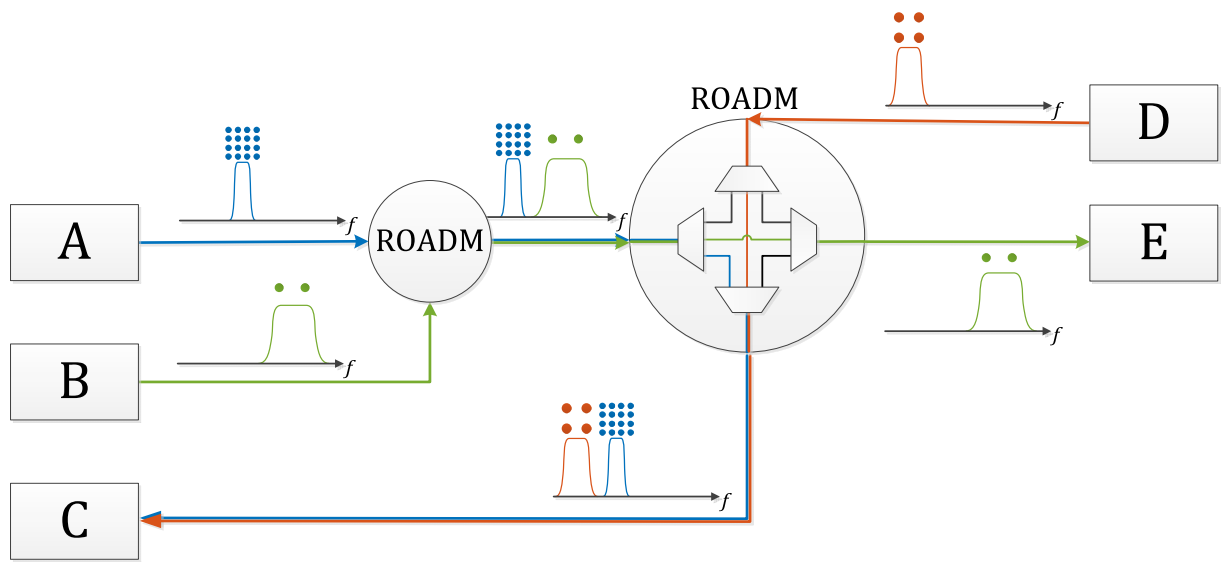

***Figure 1.1:*** *Example of a flexible optical network with data paths $\overline{AC}$ (blue), $\overline{DC}$ (red) and $\overline{BE}$ (green). Each channel is depicted by a symbolized spectrum and modulation format. The channels are routed by ROADMS.*

Another trend emerging during the last years is the transition from in-line removal of accumulated *chromatic dispersion* (CD) to CD unmanaged systems, becoming possible with the shift from direct detection schemes to optical coherent systems and the introduction of *digital signal processing* (DSP). In this systems, the accumulated CD is removed digitally at the receiver side ([6]). From this development analytical models emerged ([7]–[9]), estimating NLI fast without the need for simulations. These predictions can be used to acquire information about transmission quality and maximum transmission distance.

However, these models were originally designed for static point-to-point networks. Also, up to date research in quality-of-transmission estimation of coherent systems had its main focus on co-propagating channels. Only lately, and sparsely, the impact of channel switching is under test ([10]). The work at hand tries to fill these gaps and investigates the NLI and the reliability of the models under different conditions occurring in flexible networks: Simulations emulating the influence of accumulated chromatic dispersion, add and drop of neighboring channels, the fiber span length and channel spacing are run for this purpose. Precautions were taken to ensure the observed distortions are caused by *cross- and multi- channel interference* (XMCI) alone and other influences are removed or suppressed. These results are compared with the predictions offered by two frequency domain models, the Gaussian-noise model and the enhanced Gaussian-noise model, and interpreted by means of a time domain model, called pulse collision picture.

In connection with this work the Gaussian- and enhanced Gaussian- noise model were realized in Matlab. Further, a ROADM model and algorithms for signal evaluation were implemented. Subsequently, the full transmission simulation setup was created by means of the VPI transmission maker. To easily include the mentioned Matlab functions as well as the HHI DSP toolbox, also written in Matlab, into the setup a parser was programmed, taking care of accurate parameter transfer between Matlab and the transmission maker.

This thesis is structured in the following way:

First, some theoretical background, related but not exclusively specific to flexible optical networks, is given in **Chapter 2**. **Chapter 3** is concerned with the analytical models, their assumptions and implications. The simulative approach is then presented in **Chapter 4**, which is closely related to **Chapter5**, discussing the analysis criteria. Finally, in **Chapter 6** these criteria are used to evaluate the simulation results, which are then in turn discussed and compared to the analytical models.

# 2 Flexible Optical Networks

In fiber-optics communication systems the signal transmission can normally be divided into three main parts: transmitter, link and receiver. At the transmitter side digital data is converted into an optical signal and injected into the link. The link consists usually of a series of (single mode) fibers interrupted by amplifiers to compensate for attenuation. The optical signal is reconverted into a digital signal at the receiver. Then, by means of *digital signal processing* (DSP) the original data is recovered.

In contrast to this standard scenario, in flexible optical networks different data streams can travel simultaneously in frequency separated channels. Channels can originate from numerous locations and travel on several different paths through the network. The data inherent in a channel can also differ in its characteristics from those of other channels, for example in modulation format, symbol rate and spectral distribution (***Figure 1.1***).

These properties and their impact on the configuration of the signal are described in this chapter.

One of the main factors restricting signal transmission is the *nonlinear interference* (NLI) of other channels, called *interferer* (INT), on the *channel under test* (CUT) – *cross- and multi-channel interference* (XMCI). To grasp the origin of XMCI an understanding of wave propagation is essential, yet for the sake of brevity only aspects necessary to comprehend the results of this thesis are presented. For a more detailed picture see [11], [12] and references therein.

In literature the perturbation caused on the signal through interference is often called 'nonlinear interference noise', even though 'noise' refers to perturbations generated by random processes and the correct term would be 'nonlinear interference distortions'. This is due to the fact that, especially in the case of XMCI, where the recipient has no way of knowing the exact incidents leading to the distortions, they are often treated as noise. In order not to break with the nomenclature already established in literature, the term *nonlinear interference noise* (NLIN) is used in this thesis, although all perturbations described herein are caused by deterministic processes.

## 2.1 Optical Signals

### 2.1.1 Modulation Format

The basis of every data transmission is a discrete binary stream, which needs to be converted into a continuous optical signal. Properties of electromagnetic waves are hereby used to represent the data. This process is called modulation of the optical wave. While in early direct detection schemes only amplitude modulation was possible, with the advent of modern coherent receivers more advanced formats became feasible by also modulating the phase of the signal ([13], [14]). A modulation format is defined by its symbol alphabet, which consists of amplitude and phase values and can thus be represented by complex numbers. The representation of a modulated signal at symbol sampling instances can therefore be done as a diagram in the complex plane, called constellation diagram. The symbol alphabet of a modulation format, depicted in the same way, is also referred to as constellation. In the context of signal transmission the real and imaginary parts are usually referred to as inphase- and quadrature-component, respectively. ***Figure 2.1*** shows the constellation diagrams of two common modulation formats: *quadrature phase shift keying* (QPSK) and *quadrature amplitude modulation* (QAM). These formats are also utilized the investigation at hand.

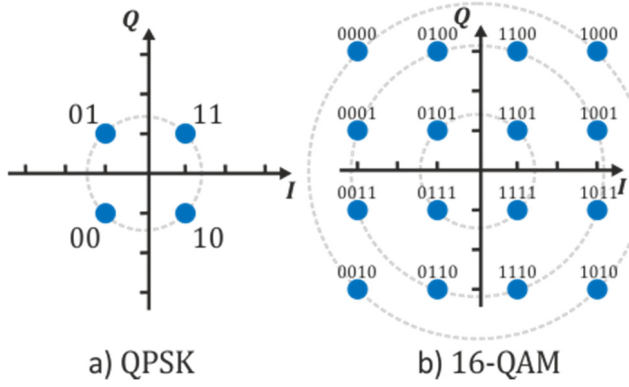
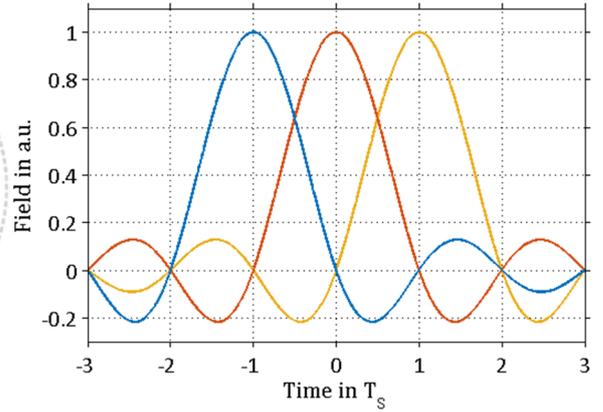

*Figure 2.1: Constellation diagrams of common modulation formats.*

*Figure 2.2: Sinc-Pulses with a symbol duration of $T_S$. At the center of each individual pulse all other pulses vanish, fulfilling Eq. (2.1).*

In general, the information is imprinted on optical pulses. This concerns the transition between symbols and is therefore not evident in the constellation diagrams but can clearly be seen in a time domain view (**Figure 2.2**).

Pulse shaping is used to suppress inter symbol interference (ISI), vanishing when the pulses satisfy the Nyquist ISI criterion

$$g_n(m \cdot T_s) = \delta_{nm}, \forall n, m \in \mathbb{Z} \tag{2.1}$$

where $g_n(t)$ is the pulse shape of the $n^{th}$ symbol in time domain, $\delta_{nm}$ is the Kronecker delta and $T_S$ is the symbol duration.

A pulse shape fulfilling this condition is the sinc-pulse shape (**Figure 2.2**):

$$g_n(t) = \frac{1}{T_S} sinc\left(\frac{t - nT_S}{T_S}\right) = \frac{\sin\left(\frac{\pi(t - nT_S)}{T_S}\right)}{\pi(t - nT_S)} \tag{2.2}$$

Its Fourier transform is a finite rectangular function of the frequency $f$ (**Figure 2.3**):

$$G_n(f) = \begin{cases} 1, & |f| \leq \frac{S_R}{2} \\ 0, & |f| > \frac{S_R}{2} \end{cases} \tag{2.3}$$

with the symbol rate $S_R = \frac{1}{T_S}$. Due to implementation penalties, e.g. *finite* length of the *impulse response* (FIR) of the filter, this pulse shape can only be approximated. By using *raised-cosine* (RC) pulses.

$$G_n^{RC}(f) = \begin{cases} 1, & |f| \leq (1-r) \cdot \frac{S_R}{2} \\ 0, & |f| \geq (1+r) \cdot \frac{S_R}{2} \\ \cos\left(\frac{\pi}{2 \cdot S_R \cdot r} \cdot \left(|f| - (1-r) \cdot \frac{S_R}{2}\right)\right)^2, & else \end{cases} \tag{2.4}$$

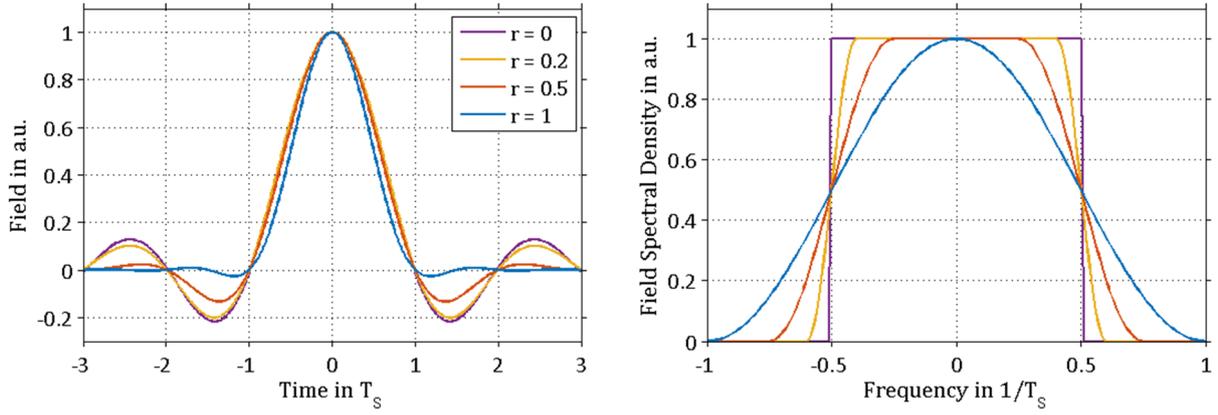

*Figure 2.3: Normalized raised-cosine functions in time domain and their Fourier transform.*

or *root-raised-cosine* (RRC) pulses,

$$G_n^{RRC}(f) = \sqrt{G_0^{RC}(f)} \tag{2.5}$$

the edges of the filter function are smoothed and the number of required FIR filter coefficients can be decreased. $r$ is a real value in $[0,1]$ and is called the roll-off of the pulse shape and defines the steepness of the filter edges. In the limit $r = 0$ the shapes of RC, RRC and sinc are identical. **Figure 2.3** shows the roll-off influence in time- and frequency domain. Pulses shaped by RRC filters with $r > 0$ do not actually fulfill the Nyquist criterion, hence they need to be filtered again by a matched filter at the receiver side, which is also of RRC shape, to acquire ISI-free character.

The sent pulse is a combination of the complex symbol $a_n$, defined by the modulation format, and pulse-shape:

$$\tilde{g}_n(t) = a_n \cdot g_0(t - nT) \tag{2.6}$$

Equally in frequency domain:

$$\tilde{G}_n(f) = a_n \cdot G_0(f) \tag{2.7}$$

The total spectrum of a channel is the sum of all individual spectral densities of the pulses transmitted in one data stream:

$$G_{Ch}(f) = \sum_n \tilde{G}_n(f) \tag{2.8}$$

### 2.1.2 Multiplexing

It was shown in the previous chapter that signals with (R)RC pulse shape and roll-off $r$ will occupy a range of $(1 + r)S_R$ of the spectrum, which is usually at the order of GHz and therefore much smaller than the bandwidth offered by optical fibers, ranging up to a few THz [15]. Consequently, it is possible to transfer more than one data stream or channel, by shifting the individual spectra to different frequency ranges. This is done by modulating the signal onto carrier waves, whose frequencies differ by more than $(1 + r)S_R$. The process is called *frequency division multiplexing* (FDM) or, more commonly used in the context of optical systems, *wavelength division multiplexing* (WDM).

The resulting spectrum is then the sum of the spectra of the individual channels:

$$G_{WDM}(f) = \sum_m G_{Ch_m}(f)$$

(2.9)

To increase the channel capacity even further, *polarization division multiplexing* (PDM) can be applied, which makes use of two orthogonally polarizations of light and thus doubles the amount of transmitted data.

## 2.2 ROADM

A *reconfigurable optical add/drop multiplexer* (ROADM) is a device that allows to selectively exchange channels in the optical domain directly, without the need to transform the signal back into digital domain, enabling fast signal splitting and rerouting.

A ROADM consists of a number of *wavelength selective switches* (WSS), optical devices capable of routing specific wavelength domains onto one of the N output ports (1xN mode). They can also be operated backwards, in Nx1 mode, filtering N input signals and combining them onto one output port. A directionless example is shown in **Figure 1.1**, the simplest ROADM example is presented in **Figure 2.4**. The latter is used in the simulations and configured in the following way: One of the ports of each device is connected and used as a pass-through whereas the other N-1 ports of the first WSS are drop ports, routing the signals onto other transmission lines or receivers if necessary. The remaining N-1 ports of the second device allow adding channels from those lines or other sources.

In principle, a WSS is a free-space optic component in which light is dispersed continuously onto a switching device – usually a LC or LC on Silicon array – mirroring the incoming wavelengths onto output ports. As the path of light is reversible, this allows for forward ('splitting') and backward ('merging') operation. The LC array allows hereby a reprogramming of the device. The WSS can be characterized by the filter functions of each individual port. Due to the components involved they usually take on shapes as shown in **Figure 2.5** ([16]).

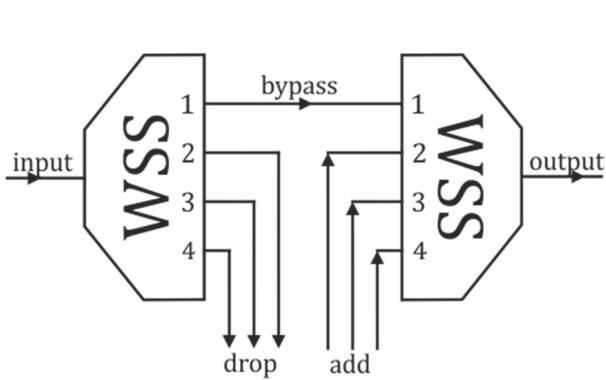
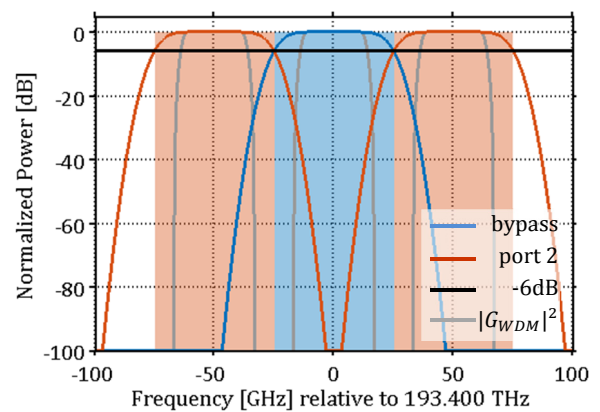

***Figure 2.4:*** *Schematic of a ROADM consisting of two 1x4 WSS, as used in the simulations. To become CDC more WSS would be needed at the add/drop ports.*

***Figure 2.5:*** *Ideal (rectangles) and analytical (respective lines) filter functions of a WSS bypassing the CUT and drop or add the INT with 50GHZ spacing. Grey: Signal PSD of 3Ch(RC, r=0.2, spacing 50GHz)*

## 2.3 Wave Propagation

It is known ([17]), that a solution for propagation of an electrical field in vacuum, according to the Maxwell equations, can be written as a homogeneous plane wave, oscillating in $\boldsymbol{e}_x, \boldsymbol{e}_y$ and travelling in $\boldsymbol{e}_z$ direction ($\boldsymbol{e}_x, \boldsymbol{e}_y, \boldsymbol{e}_z$ being spatial orthogonal unit vectors):

$$\mathbf{E}(z,t) = \mathbf{A}(z,t)\exp(i\omega t - ikz) \tag{2.10}$$

where $\mathbf{A}(z,t)$ is the complex amplitude of the electric field in $\boldsymbol{e}_x, \boldsymbol{e}_y$ including polarization of the wave, $\omega = 2\pi f$ the angular frequency of the oscillating field and $k$ is the wavenumber.

As demonstrated in **Chapter 2.1.1** the signal bandwidth is close to the symbol rate, which is usually around 20-100 GBaud. The carrier frequency $f_0$ on the other hand is matched to the low-attenuation low-dispersion range of the fiber at around 193.4THz, corresponding to a wavelength of 1550nm. This narrow-banded character leads to an only slowly varying envelope of the signal and allows to approximate Eq. (2.10) with

$$\mathbf{E}(z,t) = \mathbf{B}(z,t)\exp(i\omega_0 t - i\beta_0 z) \tag{2.11}$$

Here $B(z,t)$ is the slowly changing complex amplitude and $\beta_0 = k(\omega_0)$ the propagation constant inside the medium. The $n^{th}$ partial derivative of the real part of $\beta$ is denoted by

$$\beta_n = \left.\frac{\partial^n \Re(\beta(\omega))}{\partial \omega^n}\right|_{\omega=\omega_0} \tag{2.12}$$

whereas the imaginary part is usually considered constant in the frequency range of interest and labeled

$$\alpha \cong \alpha(\omega) = \Im(\beta(\omega)) \tag{2.13}$$

The propagation of this envelope through an optical fiber is governed by the coupled Nonlinear Schrödinger Equation (NLSE, [18]):

$$\frac{\partial \mathbf{B}(z,\tau)}{\partial z} = -\alpha \mathbf{B} + i\frac{\beta_2}{2}\frac{\partial^2}{\partial \tau^2}\mathbf{B} - i\gamma|\mathbf{B}|^2\mathbf{B} \tag{2.14}$$

The coupled NLSE can be derived from the Maxwell equations ([19]) and its parts are discussed in the subchapters to follow. There is no part depending on $\beta_1$ because the time frame was changed to $\tau = t - \beta_1 z$ which is moving with the group velocity of the pulses of the CUT along the fiber. All higher order ($n \geq 3$) derivatives of $\beta$ are assumed to be small enough to be negligible.

### 2.3.1 Attenuation

$$\frac{\partial \mathbf{B}(z,\tau)}{\partial z} = -\alpha \mathbf{B} \tag{2.15}$$

The first term on the right hand side of Eq.(2.14) concerns the optical field fiber loss $\alpha$ and its partial solution (assuming all other parameters to be zero) is

$$\mathbf{B}(z,\tau) = \mathbf{B}(0,\tau)e^{-\alpha z} \tag{2.16}$$

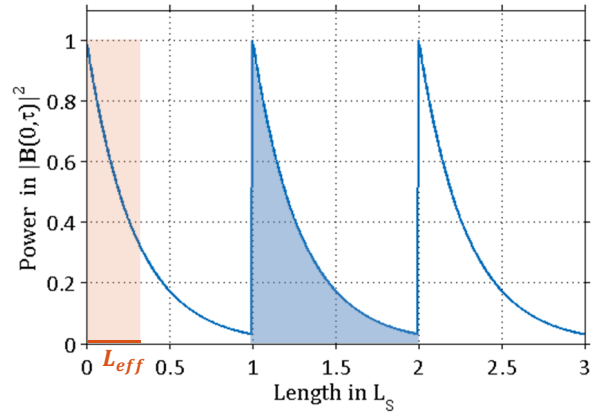

*Figure 2.6: Loss/Gain power profile over three spans of length $L_S$.*

In transmission links the signal is amplified after or during each span of length $L_S$. With lumped amplification and perfect compensation of attenuation a loss/gain power profile can be identified as

$$f(z,\tau) = e^{-2\alpha \cdot mod(z, L_s)} \tag{2.17}$$

where $mod$ is the modulo operator assuring a return to the original power after each span. A profile over three spans is shown in **Figure 2.6**. In such systems often an effective length $L_{eff}$ is defined as the length of a span with no attenuation but equal power-length-product, meaning the blue and the red area in **Figure 2.6** are of equal size. With Eq. (2.17) $L_{eff}$ becomes

$$L_{eff} = \frac{1 - e^{-2\alpha L_s}}{2\alpha} \tag{2.18}$$

### 2.3.2 Chromatic Dispersion

$$\frac{\partial \mathbf{B}(z,\tau)}{\partial z} = i\frac{\beta_2}{2}\frac{\partial^2}{\partial \tau^2}\mathbf{B} \tag{2.19}$$

*Chromatic dispersion* (CD) is described by the second term on the right hand side of Eq.(2.14) and deals with the frequency dependence of the group velocity. The resulting effect is the delay of a specific spectral component in relation to other spectral components. In time domain this leads to broadening of the pulses, while in frequency domain the corresponding outcome is a quadratic phase change over frequency (**Figure 2.7**).

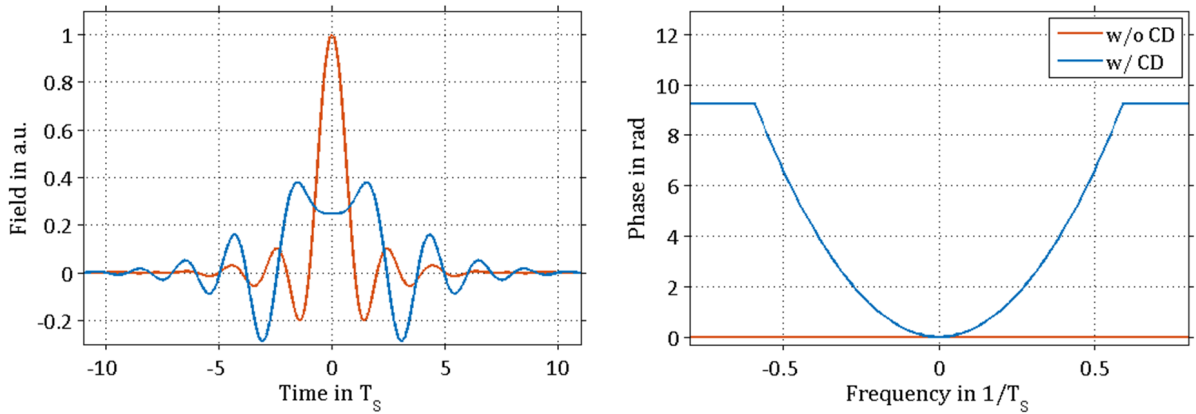

*Figure 2.7: Influence of chromatic dispersion on a raised cosine pulse (r=0.2) in time and frequency domain. The values correspond to a transmission of 80km with D=16.8ps/nm, $S_R = 28GBd$ at $f_0 = 193.4THz$. Outside the bandwidth of the pulse the amplitude of the spectral density vanishes and the phase is undefined. Here a constant value is plotted.*

### 2.3.3 Nonlinear phase modulation

$$\frac{\partial \mathbf{B}(z,\tau)}{\partial z} = -i\gamma |\mathbf{B}|^2 \mathbf{B} \tag{2.20}$$

So far, the intensity dependence of the refractive index, the Kerr effect (Eq.(2.22)), was not mentioned. The remaining term of the coupled NLSE, Eq.(2.20), is a result of this effect and therefore also sensitive to the intensity of the envelope signal. The parameter $\gamma$ accounts for the fiber properties.

$$\gamma = \frac{2\pi \omega_0 n_2}{c_0 A_{eff}} \tag{2.21}$$

Here, $A_{eff}$ is the fiber effective area, $c_0$ the speed of light in vacuum and $n_2$ the nonlinear part of the refractive index

$$n = n_1 + n_2 I \tag{2.22}$$

Kerr nonlinearities lead to a set of phenomena called *Self-Phase Modulation* (SPM), *Cross-Phase Modulation* (XPM) and *Four-Wave-Mixing* (FWM), depending on the number of frequencies involved. In its non-degenerate form three frequency components of the optical field with frequencies $f_1, f_2, f_3$, propagating simultaneously through a fiber, generate a fourth field at frequency $f$. Due to the necessity of phase matching the four frequencies fulfill the condition

$$f = f_1 + f_2 - f_3 \tag{2.23}$$

In the context of multi-channel fiber links another vocabulary is used to describe the same effects, but distinguishing not the number of frequency components but the number of channels involved in causing the NLI at the CUT. The terms *Self-Channel Interference* (SCI), *Cross-Channel Interference* (XCI) and *Multi-Channel Interference* (MCI) are employed: SCI is caused by the CUT itself, XCI by the beating of the CUT and any single *interfering channel* (INT) and MCI by the beating of three different channels ([20]), which is shown in **Figure 2.8**. Here, the red and blue curves represent the $G_{WDM}$ spectrum.

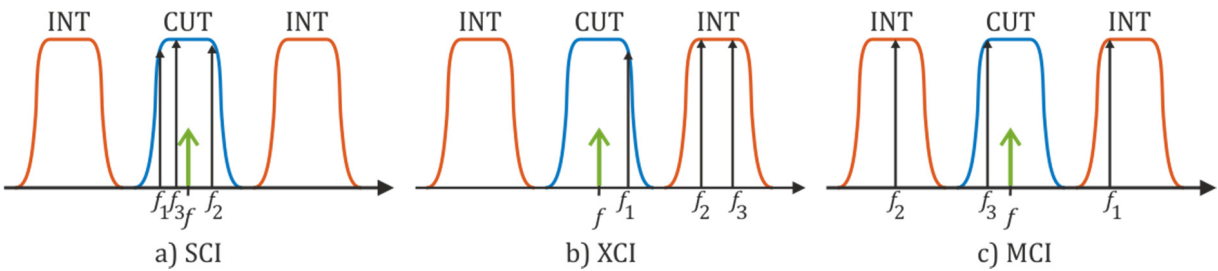

**Figure 2.8:** *During propagation the spectrum at frequencies $f_1, f_2, f_3$ causes nonlinear interference at $f = f_1 + f_2 - f_3$.*

# 3 Analytical Transmission Models

The equations presented in the previous section describe the signal development during propagation through the link. In general, there are two methods to find solutions for these equations: numerical simulation and analytical approximation. In this chapter, three state of the art models are presented, which were derived by means of the latter approach from the NLSE. They do not yield solutions for the NLSE per se, but properties extracted from these solutions.

Two of them, the *Gaussian Noise Model* (GN-model) and the *Enhanced Gaussian Noise Model* (EGN-model), are used to yield quantitative results. They are actually only semi-analytical in a sense, that they are developed analytically but still need numerical solving of integrals. The third approach, the *pulse collision picture* (PC-picture), is here used only for qualitative understanding of the process of NLI. It is presented before the EGN-model, as the ideas of the EGN-model are partly based on this picture.

In the context of the following chapters the expression 'orientation' is used to describe the properties of the spatial distribution of the received symbols in the I,Q-plane as well as their relation to each other. Also, if a symbol is described as perpendicular to another symbol, this means that their position vectors in the I,Q-plane are orthogonal.

## 3.1 Gaussian Noise Model

The GN-model is a FWM-type perturbation model for estimating NLI in a link which relies heavily on the distortion of the signal by chromatic dispersion.

FWM-type models are frequency-domain based models, expressing the beating of spectral components during propagation analytically in a fashion similar to FWM (**Chapter 2.3.3**). Already in 1993 a FWM approach was developed by the group of Petermann [21], and used 10 years later to model the *power spectral density* (PSD) of the NLI of a WDM system [22]. Tang et al. approached the topic by means of Volterra-Series in 2002 [23] and arrived at similar conclusions concerning the nature of the NLI. None of those stirred much interest at the time for they were mainly concerned with dispersion unmanaged systems, but dispersion compensated links were still common.

Only recently, in 2011 – the amount of dispersion uncompensated transmission systems now growing - the group of Poggiolini developed an equivalent model [7] and named it GN-model in 2012 [24], due to the assumptions involved. It was since topic of several follow up papers ([20], [25], [26],[27]) and enhancements ([8], [28]–[30]).

In the GN-Model three assumptions are made to approximately solve the coupled NLSE:

1. The distortion caused by NLI on the signal is small with respect to the power of the signal itself. This assumption is the basis for every perturbation approach.
2. Chromatic dispersion on the signal pulses leads to Gaussian signal statistics.
3. The distortion from NLI on the signal has Gaussian noise character

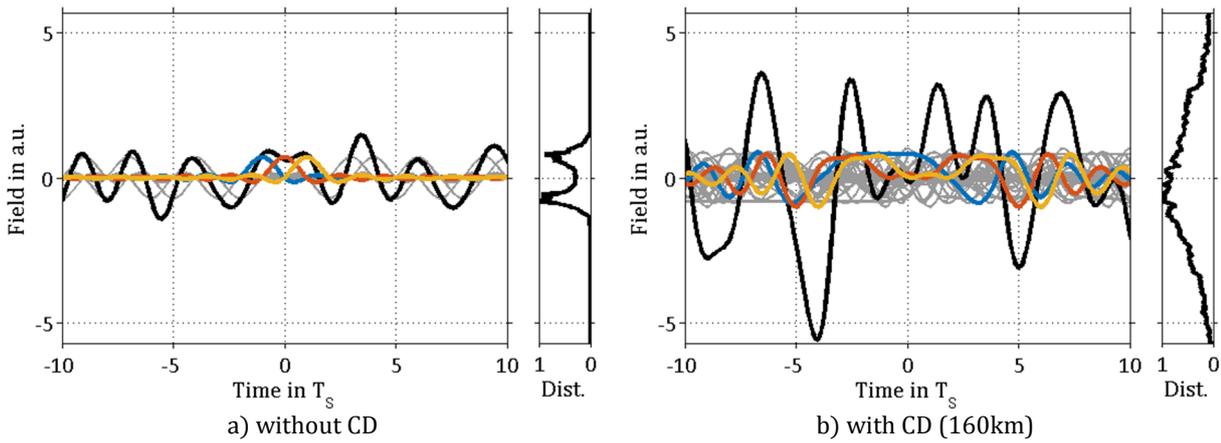

***Figure 3.1:*** *The sum (black) of the QPSK mapped pulses (gray & color) of a signal of 400 sinc-pulses with 28GBd at 193.4THz, travelling a fiber with dispersion coeff. of 16.8 ps/nm/km (zero attenuation) and periodic boundary conditions. The normalized(to the maximum value for each plot individually) distribution is shown by the histogram to the right of the plots.*

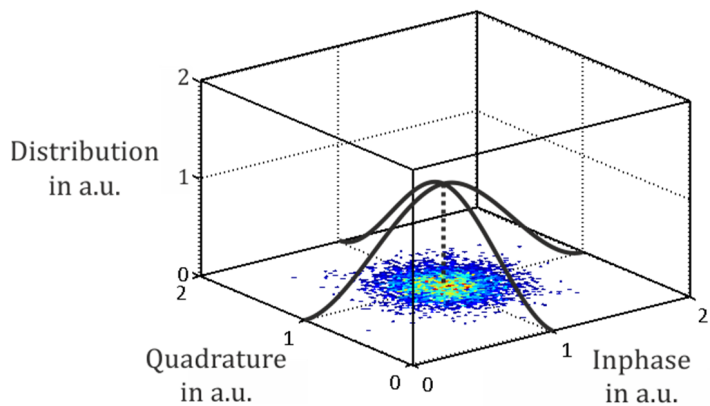

***Figure 3.2:*** *The disturbance induced by NLI in the field is Gaussian distributed in I and Q around the constellation point (here at (1,1)) according to the third assumption of the GN-model. While the data was taken from a simulation acc. to chapter 4 the distribution shown by the black curve is only an illustration.*

***Figure 3.1*** shows the idea on which the second assumption is based: While the field of the signal peaks at the amplitude of the symbols (here two peaks because QPSK modulated symbols are shown) and falls off exponentially from this points (a)), accumulating chromatic dispersion by propagation along the fiber causes the field distribution to acquire another shape, which is assumed to be Gaussian (b)). Because of energy conservation the sum of the histogram bins of a) and b) is identical, therefore the non-normalized maximum of the density of b) is actually much smaller than that of a).

***Figure 3.2*** shows a distribution of received symbols distorted by NLI. This distribution around a constellation point, here at (1,1) in I and Q, is also presumed to be Gaussian, according to the third assumption.

While here only the general gist of the two assumptions is shown, a more detailed statistical analysis can be found in the appendices of [24].

The twofold gaussianity of the second and the third assumption is the reason for naming this model. Their validity was demonstrated in [24] but then again criticized in [31], which led to the follow up models presented in **Chapter 3.2** and **Chapter 3.3**. Nonetheless, for a lot of cases, especially for modulation formats with multiple amplitude levels like higher order QAM formats as well as lumped amplification transmission schemes with transmission lengths of the order of a few hundred kilometers and longer, the assumptions hold and the GN-model yields good results.

The main formula of the GN-model, Eq. (3.1), can be derived by applying a perturbative approach to the NLSE Eq.(2.14) and using the three assumptions ([25]).

$$|G_{NLI}^{N_S}(f)|^2 = \frac{16}{27} \cdot \int_{-\infty}^{\infty}\int_{-\infty}^{\infty} |G_{WDM}(f_1)|^2 \cdot |G_{WDM}(f_2)|^2 \cdot |G_{WDM}(f_1+f_2-f)|^2 \cdot |\mu(f_1,f_2,f)|^2 df_1 df_2$$
(3.1)

The interference from both polarizations is here already incorporated. As before $G_{WDM}(f)$ is the field spectral density of the WDM channels. The output $|G_{NLI}^{N_S}(f)|^2$, the *power spectral density* (PSD) of the NLI generated by span $N_S$ at frequency $f$, is produced by the PSD of the signal at frequencies $f_1, f_2$ and $f_3 = f_1 + f_2 - f$, therefore it is easy to recognize the model as FWM-like.

$\mu(f_1, f_2, f)$ weights the formation of NLI depending on link parameter, including a factor responsible for the efficiency of the FWM beating ($\zeta(f_1, f_2, f)$) and one for the coherence of the perturbation in consecutive spans of the link ($\nu(f_1, f_2, f)$). For lumped amplification the factors take the following form:

$$\mu(f_1, f_2, f) = \zeta(f_1, f_2, f) \cdot \nu(f_1, f_2, f)$$
(3.2)

$$\zeta(f_1, f_2, f) = \gamma \frac{1 - e^{-2\alpha L_s} e^{i4\pi^2 \beta_2 (f_1-f)(f_2-f) L_s}}{2\alpha - i4\pi^2 \beta_2 (f_1-f)(f_2-f)}$$
(3.3)

$$\nu(f_1, f_2, f) = \frac{\sin(2\pi^2 \beta_2 (f_1-f)(f_2-f) N_s L_s)}{\sin(2\pi^2 \beta_2 (f_1-f)(f_2-f) L_s)} e^{i2\pi^2 \beta_2 (f_1-f)(f_2-f)(N_s-1) L_s}$$
(3.4)

From Eq.(3.1) it is also clear that any phase information of the signal is ignored, due to the norm-operations. The GN-model takes therefore additional accumulated CD on the channels, group delay and phase due to modulation format not into account and is instead based solely on the PSD of the signal.

The authors of [7] claim that, in accordance with the second assumption, CD makes the signal 'more Gaussian' during propagation, diminishing the discrepancy between the model and simulations of non-Gaussian channels, which is most pronounced during the first spans of the transmission ([8]). This makes the model better for longer transmission length or pre-dispersed channels ([27]).

Implementation of the GN-model was also part of the work done in context of this thesis. Calculation is about 10 times faster than obtaining the same results by simulation.

In the GN-model there is no discrimination between SCI, XCI and MCI contributions. In order to retrieve XMCI, a full setup including all INT and a setup consisting only of the CUT need to be realized and then subtracted from each other.

## 3.2 Pulse-Collision Picture

In [32] a modulation format dependence of the NLI was predicted, which is ignored in the GN-model by assuming all channels to be of Gaussian nature. The calculations were picked up again in [31]: Using first order perturbation theory plus the assumption of mainly XCI contribution, stated in [5], properties of NLI in time- and frequency-domain were developed, still supporting the dependence on modulation format.

To understand this dependence a new picture was presented in [9]. In contrast to the GN-model this model is time-domain based and hence it deals with the signal representation in terms of the field instead its spectrum. The general idea is, that the interactions between the pulses making up the signal can be approximated as perturbations caused by 'collisions' of the pulses - thus leading to the name 'pulse-collision picture'.

The influence of chromatic dispersion is now included in the pulse shape, Eq. (2.2), which becomes therefore dependent on the location of the pulse along the transmission direction $z$.

$$g_n(z,\tau) = e^{-i\frac{\beta_2}{2}\cdot z \cdot \frac{\partial^2}{\partial \tau^2}} g_n(\tau) \tag{3.5}$$

With $a_n$ the symbols of the CUT, $b_{n,w}$ the symbols of the $w^{th}$ neighboring channel and $\Omega_w$ their angular frequency distance to the center channel, the field envelope at $z$ becomes

$$B^{(0)}(z,\tau) = \sum_n a_n\, g_0(z,\tau - nT_S) + \sum_w \sum_n b_{n,w}\, g_0(z,\tau - nT_S - \beta_2\Omega_w z) \tag{3.6}$$

assuming identical pulse shaping of all channels. In this context attenuation is compensated ideally at the receiver side. To take account of its influence on the non-linear term, the loss/gain profile $f(z)$ (Eq.(2.17)) is included in the respective term in the NLSE. Calculating the first order perturbation $B^{(1)}$ on the linear solution $B^{(0)}$ becomes then

$$\frac{\partial \mathbf{B}^{(1)}(z,\tau)}{\partial z} = i\frac{\beta_2}{2}\frac{\partial^2}{\partial \tau^2}\mathbf{B}^{(1)}(z,\tau) - i\gamma f(z)\big|\mathbf{B}^{(0)}(z,\tau)\big|^2 \mathbf{B}^{(0)}(z,\tau) \tag{3.7}$$

with the solution at the receiver position L

$$\mathbf{B}^{(1)}(L,\tau) = i\gamma \int_0^L e^{-i\frac{\beta_2}{2}\cdot(L-z)\cdot\frac{\partial^2}{\partial \tau^2}} f(z)\big|\mathbf{B}^{(0)}(z,\tau)\big|^2 \mathbf{B}^{(0)}(z,\tau) dz \tag{3.8}$$

Additionally, the signal is filtered at the receiver with a matched filter, the complex conjugated pulse-shape of the symbol $g_n^*(L,\tau)$. Inspecting only symbol $a_0$, without loss of generality, the received perturbed symbol $\hat{a}_0$ will become

$$\hat{a}_0 = a_0 + \Delta a_0 \tag{3.9}$$

$$\begin{aligned}\Delta a_0 &= \int_{-\infty}^{\infty} \mathbf{B}^{(1)}(L,\tau)\, g_0^*(L,\tau)\mathrm{d}\tau \\ &= i\gamma \int_0^L f(z) \int_{-\infty}^{\infty} g_0^*(L,\tau)\big|\mathbf{B}^{(0)}(z,\tau)\big|^2 \mathbf{B}^{(0)}(z,\tau)\, \mathrm{d}\tau\, \mathrm{d}z\end{aligned} \tag{3.10}$$

Inserting Eq. (3.6) will now yield a formula accounting for SCI, XCI and MCI, yet the authors of the pulse-collision picture argued, that SCI can be compensated and MCI is negligible. Therefore all terms involving only the CUT and all terms involving more than one INT were dropped.

$$\Delta a_0^{XCI} = \sum_w \Delta a_{0,w}^{XCI} = \sum_w 2i\gamma \sum_{h,k,m} a_h b_{k,w}^* b_{m,w} X_{hkm,w} \tag{3.11}$$

3 Analytical Transmission Models

$$X_{hkm,w} = \int_0^L f(z) \int_{-\infty}^{\infty} g_0^*(z,\tau) \, g_0(z, \tau - hT_S) \, g_0^*(z, \tau - kT_S - \beta_2 \Omega_w z) \, g_0(z, \tau - mT_S$$
$$- \beta_2 \Omega_w z) \, d\tau \, dz \qquad (3.12)$$

Since the influence of the channels simply add up, in the following only one channel will be examined and the index $w$ will be dropped. The resulting perturbation can be divided into four parts, depending on the degeneracy of the involved pulses:

$$\Delta a_0^{XCI} = 2i\gamma \left( \sum_{h=0, k=m} a_0 |b_m|^2 X_{0mm} + \sum_{h=0, k \neq m} a_0 b_k^* b_m X_{0km} \right.$$
$$\left. + \sum_{h \neq 0, k=m} a_h |b_m|^2 X_{hmm} + \sum_{h \neq 0, k \neq m} a_h b_k^* b_m X_{hkm} \right) \qquad (3.13)$$

Term $h = 0, k = m$: Here, only two pulses are involved, one of the CUT and one of the INT, therefore it is called two-pulse collision term. The overlap of the $g_0$ in Eq.(3.12) is maximized and hence this term yields the biggest contributions. Also, since there are only $|g_0|^2$ involved $X_{0mm}$ is real, as is $|b_m|^2$ and therefore its contribution is always proportional to $ia_0$. It is therefore perpendicular to $a_0$ and has approximately the character of a phase rotation. The variance over all symbols of this term, indicating its contribution to the NLIN, is proportional to $E[|b_m|^4] - E[|b_m|^2]^2$. This difference, being zero when all symbols have the same amplitude like QPSK and of finite value if there are multiple amplitude levels, like 16QAM, results in modulation format dependence. For QPSK INT two pulse collisions will therefore result only in phase rotations of the CUT in its entirety and will not contribute to noise.

Term $h = 0, k \neq m$: This term is called type A three-pulse collision term, as there are three pulses involved, but to distinguish it from the term $h \neq 0, k = m$. Examining the sum $a_0 b_k^* b_m X_{0km} + a_0 b_m^* b_k X_{0mk} = a_0 b_k^* b_m X_{0km} + a_0 b_m^* b_k X_{0km}^* = 2a_0 \Re(b_k^* b_m X_{0km})$ it is obvious that, as before, the perturbation from this term is perpendicular to $a_0$ and results in phase rotations. The variance on the other hand is now proportional to $E[|b_m|^2]E[|b_k|^2] - |E[b_m]|^2|E[b_k]|^2 = E[|b_m|^2]^2$ and thus depends on the power of the modulation only.

Term $h \neq 0, k = m$: Like before three pulses are involved, only this time two of the CUT and one of the INT. In the advance of this thesis the effect of this term will be called type B three-pulse collisions. Because now there is also $a_h$ involved, there is no fixed orientation relation to $a_0$ – the resulting perturbation is randomly circular distributed. Due to this distribution this noise will be called Gaussian noise or circular noise in the following chapters. Similar to two-pulse collisions the variance of the noise depends again on $E[|b_m|^4] - E[|b_m|^2]^2$, on the second and fourth moment of the INTs modulation format.

Term $h \neq 0, k \neq m$: This term concerns the interference of four different pulses. It has neither an orientation relation to $a_0$, the resulting NLIN is therefore also circular distributed, nor a dependence on the modulation format. This term includes the most indices and has therefore the most summands. The overlap between the pulses on the other hand is the least and the $X_{hkm}$ usually the smallest.

The assessment about the magnitude of the $X_{hkm}$ is only valid if the collisions are complete, i.e., the pulses have fully passed each other. ***Figure 3.3*** shows the development of the coefficients during collision for the different types in terms of current NLIN contribution (blue) and accumulated NLIN (green). At the left of the graphs the collision starts and is completed at the distance at the right side of the plot. It can be seen that the contribution from two-pulse collisions is largest in case of complete collisions, as the built up of NLIN is monotonously increasing. In case of the other pulse collisions on the other hand, an interruption of the process at an intermediate state can result in bigger NLIN than a complete collisions. In transmission setups using lumped amplification the completeness of the collisions is not guaranteed: the collisions are interrupted by amplifiers at random collision stages. The contributions from two-pulse collisions are reduced by this effect, while the influence of the other types is enhanced. Especially NLI originating from four-pulse collisions will rise due to the large relative number of those collisions.

The incompleteness is boosted even further when accumulated CD is taken into consideration. As shown in **Figure 2.7** the pulse width is increased by CD and hence the passing time of two pulses is prolonged. In the eyes of the authors of [9] the shift to incomplete collisions is the true cause for better agreement with the GN-model with longer transmission distance or when increasing the accumulated CD.

The pulse collision picture presented here does only account for signals with one polarization and is therefore only to a limited extend suitable for PDM signals.

The properties of the collision contributions are summarized in **Table 3.1**.

|  | **Two-Pulse Collisions** | **Three-Pulse Collisions** |  | **Four-Pulse Collisions** |
|---|---|---|---|---|
|  |  | Type A | Type B |  |
|  | $h = 0, k = m$ | $h = 0, k \neq m$ | $h \neq 0, k = m$ | $h \neq 0, k \neq m$ |
| **Interference Character** | Phase | Phase | Gaussian | Gaussian |
| **Mod. Format dependence** | Yes | No | Yes | No |
| **Most contributing collision type** | Complete | Incomplete | Incomplete | Incomplete |

*Table 3.1: Summary of pulse collision types and their properties.*

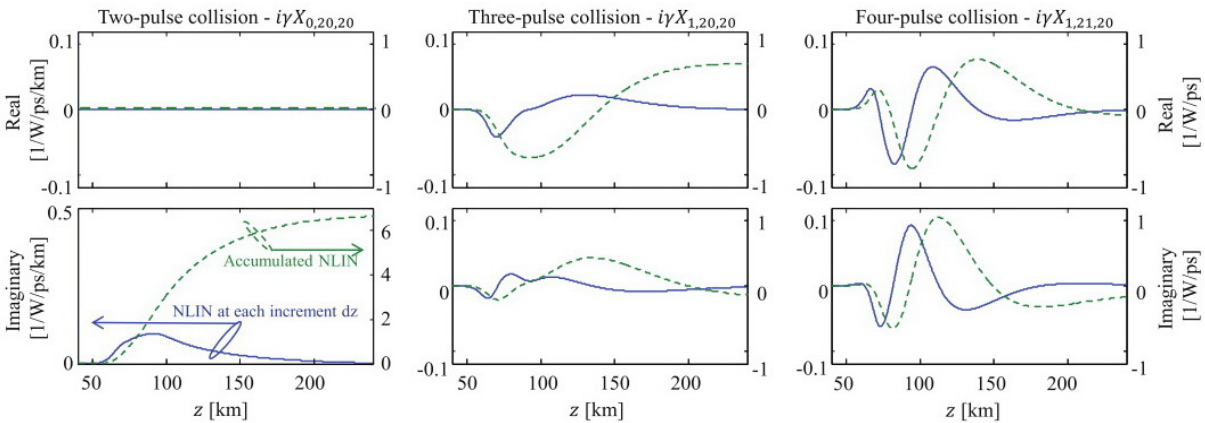

*Figure 3.3: Accumulation of NLI during pulse collision. Z is the distance the pulses propagate while passing each other, it depicts therefore also a progression in time. Examples shown for RRC pulses wit r=0.2, 32GBaud, $\beta_2 = 21 ps^2/km$, $\gamma = 1.3 \frac{1}{W \cdot km}$, 50GHz channel spacing and zero attenuation. This figure was taken from [9].*

## 3.3 Enhanced Gaussian Noise Model

The last model in this comparison is an advancement of the GN-model by the same authors. Based on the work about modulation format dependence in [5] and inspired by the development done in [31] the assumption of signal gaussianity was set aside and a new model emerged: the EGN-model. In contrast to [31],[5] and the PC-picture [9] not only XCI contributions but also SCI and MCI were included. On the other hand, perfect rectangular spectra were assumed.

The resulting equations can be divided into parts independent of modulation format, resembling the GN-model, and correction terms depending on modulation.

$$\left|G_{NLI}^{EGN}(f)\right|^2 = \left|G_{NLI}^{GN}(f)\right|^2 + \left|G_{NLI}^{corr}(f)\right|^2 \tag{3.14}$$

All terms in Eq. (3.14) can further be split up into three parts:

$$\left|G_{NLI}(f)\right|^2 = \left|G_{SCI}(f)\right|^2 + \left|G_{XCI}(f)\right|^2 + \left|G_{MCI}(f)\right|^2 \tag{3.15}$$

While these parts are of identical structure in case of the GN-model, the EGN model includes up to $6^{th}$ order moments. The values of the moments differ depending on the combination of channels involved, similar to the PC-picture, causing the terms to differ in their coefficients and configuration. Therefore, the equation breaks down into many terms; one of the correction formulas for XCI contribution, for example, involving the CUT and one INT takes the following form:

$$\left|G_{XCI_{12}}^{corr}(f)\right|^2 = \frac{80}{81}\Phi_b S_R^2 P_{CUT} P_{INT}^2 \int_{-\infty}^{\infty} |G_{CUT}(f_1)|^2 \left|\int_{-\infty}^{\infty} G_{INT}(f_2) \right.$$
$$\left. \cdot G_{INT}^*(f_1 + f_2 - f) \ \mu(f_1, f_2, f) df_2 \right|^2 df_1 \tag{3.16}$$

$G(f)$ are normalized channel spectra, such that $\int |G(f)|^2 df = 1/S_R$, which accounts for the $S_R^2$ in the coefficient. $P_{CUT}$ and $P_{INT}$ are the powers of the CUT and INT respectively. $\Phi_b = E[|b|^4]/E[|b|^2]^2 - 2$ takes care of the modulation format dependence. For Gaussian distributed amplitudes $\Phi_b = 0$ and the whole term vanishes. Also, it is obvious that here the phase of the signal is included.

There are more terms, calculating the overlap of the two channels differently. For example $G_{CUT}$ could be integrated over $f_2$ and $G_{INT}$ could be in the outer integral. Each combination results in different moments, in this case resulting in the coefficient 80/81, and each combination has to be calculated to retrieve $|G_{NLI}^{corr}(f)|^2$.

The big number of terms increases the accuracy of the model, yet also increases its complexity. The integrals of each channel combination need to be calculated individually instead of one single integral, doubling computation time. The EGN model was also implemented in the course of this thesis, yet the implementation it does not support adding accumulated CD on the channels.

# 4 Numerical Simulation Setup

Another approach to solve the NLSE and estimate the NLI of an optical transmission system is to simulate the system. In contrast to the analytical models, calculating only the PSD of the distortion, the propagation of the signal itself is modeled in a simulation - from its creation to its conversion back into data.

In simulations, the same *digital signal processing* (DSP) can be used which would be applied to real signals. Furthermore, when interested in specific influences on the signal, simulations allow turning off other impacts, which might mask the influence one is interested in. In actual systems impairment of the signal occurs at all stages of the transmission, due to component imperfections and physical limitations. XMCI, however, takes place only during propagation through the link. As a result most components are assumed to be perfect.

In this section the tools used for simulation and DSP are presented and their algorithms shortly discussed. All simulations were conducted with periodic boundary conditions.

## 4.1 VPI Transmission Maker

An overview over the simulation setup can be seen in ***Figure 4.3***, the corresponding parameters can be found in ***Table 4.3***: At the transmitter DSP (**Chapter 4.2.1**) an electrical signal is created, which is then modulated onto an optical wave of constant frequency and power emitted by a laser. To avoid numerical artifacts the signal is afterwards randomly rotated and delayed, which would also be the case in real systems, due to the transmission paths through the components. The signal is then multiplexed with eight similarly created optical waves of different frequency, whose modulation format, frequency shift and accumulated chromatic dispersion are defined according to the current parameters under investigation (see **Chapter 6**). Before entering the loop, the signal is amplified. The loop itself consists of a component simulating the transmission through the fiber, a ROADM model and an amplifier to compensate for fiber attenuation. One cycle of the loop therefore simulates a span of the link. After each loop iteration the signal is evaluated at the receiver, but also a duplicate is sent back to the beginning of the loop simulating transmission through the next span. At the receiver side the conversion from optical to electrical domain is done by a coherent frontend, consisting of a polarization splitter and a 90° hybrid and four photodiodes for each polarization. The final step of transmission is then again assumed by DSP (**Chapter 4.2.2**).

For this thesis the graphical simulation tool VPI transmission maker was used for the optical part of the simulation. It was chosen for its variety of implemented optical components and its advanced fiber model.

The latter is based on the 'split-step Fourier method' ([12]) for solving the NLSE: The operators of the NLSE are divided into an operator $\widehat{D}$ taking care of the linear part and an operator $\widehat{N}$ accounting for the non-linear part. Eq. (2.14) thus becomes

$$\frac{\partial \mathbf{B}(z,\tau)}{\partial z} = (\widehat{D} + \widehat{N})\mathbf{B}(z,\tau) \qquad (4.1)$$

The itself fiber is divided into small steps of length $dz$, depending on the desired maximum phase change of the signal, over which the non-linear influences and the linear influences are assumed to operate independently – their operators approximately commutate: $[\widehat{D}, \widehat{N}] = \sigma(dz^2)$. In the symmetric implementation of the algorithm, which was used for the simulations, $\widehat{D}$ drives the signal in frequency domain over the length of $dz/2$, which is simply a multiplication due to the linearity of the operator. The signal is then converted into time domain by FFT and the wave equation $\frac{\partial \mathbf{B}(z,\tau)}{\partial z} = \widehat{N}\,\mathbf{B}(z,\tau)$ solved by numerical integration. After a transformation back into frequency domain, the second linear half of $dz$ is computed.

It is important to notice that in order to calculate the influence of birefringence, random orientation and strength of the birefringence eigenstates is chosen when going to the next step. This is in agreement with another method, called the 'coarse step method' usually applied in simulations with *polarization mode dispersion* (PMD, [11]). To disregard any effects that may be caused by choosing a specific combination of PMD matrices five different realizations of the channel are simulated, each specified by a fixed random seed. The results are then averaged over these five realizations. Additionally, whenever signals are randomly delayed and rotated in polarization to avoid numerical artifacts, the random values are also fixed for each realization.

There was no ROADM implementation fulfilling the requirements of this investigation available in the transmission maker, hence a MATLAB algorithm was implemented according to **Chapter 2.2**. While this program is capable of reproducing the filter shapes shown there, the filters were assumed to be perfectly rectangular to separate the influence of filter penalties from the influence of NLI.

A final remark about the EDFAs: In all simulations the creation of ASE-noise was omitted. Some of the simulations were repeated including ASE generation by the EDFAs, yet the results changed only by the added amount of circular noise power and the additional noise had no other consequences. It was also verified by direct comparison, that ASE-noise is unaffected by non-linear influences due to its small power, similar to the case below, showing that NLI induced perturbations are not contributing to NLI themselves.

## 4.2 Digital Signal Processing

The DSP is done in MATLAB with the help of the DSP Toolbox from HHI with slightly changed algorithms adapted to the needs of the analysis and an additional function for distortion estimation and characterization.

### 4.2.1 Transmitter DSP

*Figure 4.1* shows the workflow of the DSP at the transmitter side. First the binary data is generated by a random generator and afterwards modulated to 16QAM. Then, a header is inserted, for the equalization at the receiver side will be data aided: the filter coefficients of the *frequency domain equalizer* (FDE) are calculated from received training symbols, which are known by the equalizer. Here, *constant amplitude zero autocorrelation* (CAZAC) training symbols are used. The modulated data stream is afterwards imposed onto pulses at 2 samples per symbol. The modelling of a *digital-to-analog converter* (DAC) was omitted, emulating a perfect DAC, with optimal filtering and no quantization of the signal. Finally, the signal is passed to the VPI transmission maker and converted into an optical signal, see **Chapter 4.1**.

### 4.2.2 Receiver DSP

*Figure 4.2* summarizes the DSP at the receiver side: The CUT is extracted at the beginning of the DSP by a rectangular filter of width matching the channel spacing. In case of a real signal the CUT would be extracted before coherent detection, due to the limited bandwidth of the components, and converted back to digital domain by an *analog-to-digital converter* (ADC). In the simulation there is no bandwidth limitation and the filtering can be done in electrical domain. Analogous to the DAC the ADC was assumed to be ideal.

During the next steps a few precautions were taken to ensure the extraction of the influence of only XMCI on NLIN. First of all, to remove NLIN caused by SCI and to remove CD, the CUT is digitally propagated back along the link by a similar algorithm ([33]) as used by the VPI transmission maker but with inverted parameters. As only one channel is propagated, this can be done with only small computational effort. This method is thoroughly investigated and will be implemented in future commercial systems ([34],[35], [36]). One might argue that now the NLIN is back-propagated to the beginning of the link, too. Yet, the power of the NLIN is very small compared to the signal power, therefore the perturbation caused by NLIN should be insignificant. This property was verified by subtracting the NLIN of a single channel simulation from the NLIN of a WDM simulation with identical CUT and thus removing the SCI influence.

From the received training symbols, the channel transfer matrix is estimated and the filter coefficients for the *frequency domain equalizer* (FDE) calculated. As the distortions on the trainings symbols also include non-linear influences the linear FIR filter becomes less exact, leading to additional penalties on the received symbols. To assure an optimal equalization, a single channel simulation with zero fiber non-linearities was conducted beforehand. The FDE filter coefficients of each realization and transmission distance were saved and later used to remove the link impairments from the signal. This is a procedure only applicable in simulations as the transfer function of the link changes constantly in actual systems.

After the FDE the signal is sampled down to one sample per symbol and normalized.

The final two steps are specific to this estimation scheme. In actual data transmission systems the symbols of the modulation format would be assigned to the received symbols ('decision') and the data would be recovered. Here, the knowledge of transmitted data is used to first restore the phase of the signal and then for distortion characterization, the latter being explained in detail in **Chapter 5**.

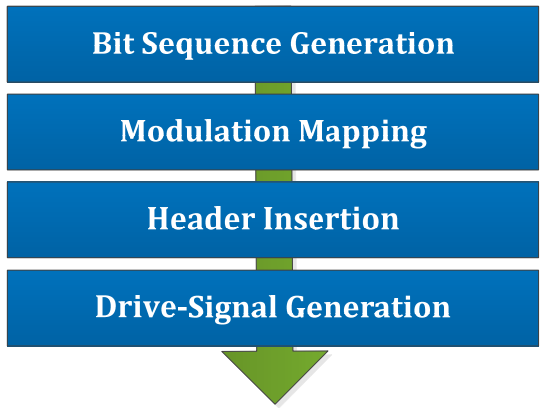

*Figure 4.1*: Transmitter DSP Flowchart

| Transmitter DSP: | |
|---|---|
| Symbol Rate | $28 GBaud$ |
| Number of Symbols | $2^{16}$ |
| Training Sequence | $CAZAC$ |
| Train. Seq. Length | $8 \cdot 64 \; (= \; 2^9) \; Symb$ |
| Polarizations | $2$ |
| Modulation CUT | $16QAM \; (4Bit/Sym)$ |
| Modulation INT | $16QAM$ |
|  | $[QPSK, Gaussian]$ |
| Channels | $9$ |
| Channel Spacing | $37,5 GHz$ |
|  | $[50 GHz, 62,5 GHz]$ |
| Pulse Shape | root raised cosine |
| Roll-Off | $0.2$ |
| Samplerate | $2 \; samples/symbol$ |

*Table 4.1*: Simulation parameter of the transmitter DSP. Values in [ ] are only used if explicitly stated in the text.

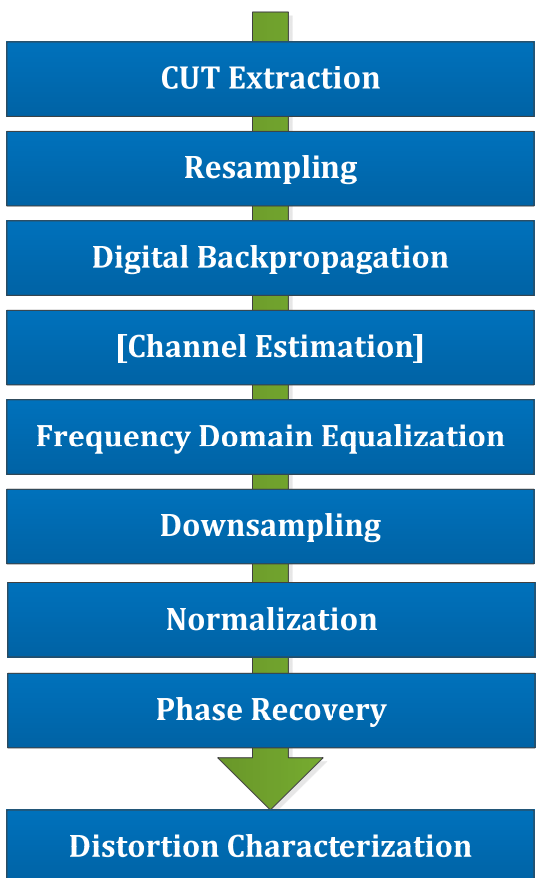

*Figure 4.2*: Receiver DSP Flowchart

| Receiver DSP: | |
|---|---|
| Filter Bandwidth | channel spacing |
| Filter Shape | rectangular |
| Samplerate | $2 \; samples/symbol$ |
| FDE Criterion | loaded coefficients |
|  | [zero forcing] |
| FDE Tap Size | $128$ |

*Table 4.2*: Simulation parameter of the receiver DSP. Values in [ ] are only used if explicitly stated in the text.

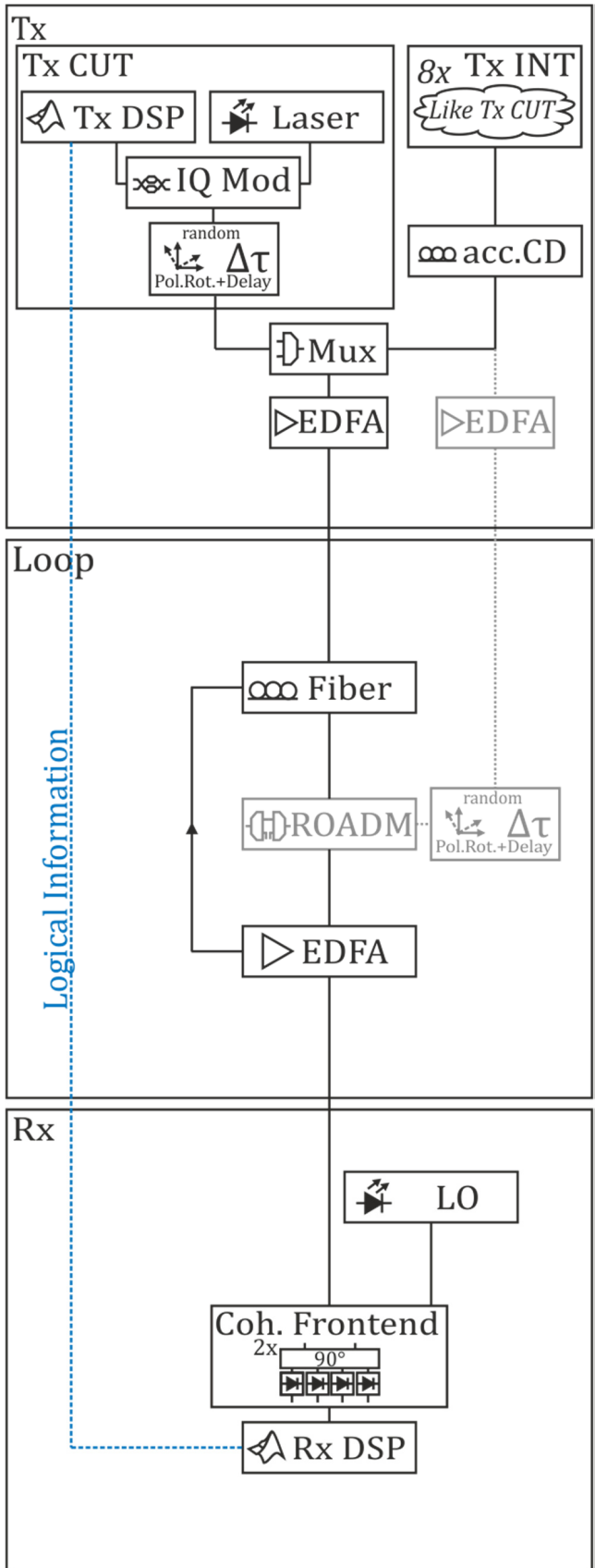

| Laser: | |
|---|---|
| Center Freq. CUT | 193.4 $THz$ |
| Center Freq. INT | " +channel spacing |
| Linewidth | none |
| **IQ-Modulator** | |
| | ideal |
| **Acc. Dispersion** | |
| Pre-Imposed | 0 $ps/nm$, |
| Acc. Dispersion | [13000 $ps/nm$] |
| **Multiplexer** | |
| | ideal |
| **Amplifier (Tx)** | |
| Output Power | 3dBm per channel |
| Noise | none |
| **Fiber** | |
| Length | 80 $km$, [40 $km$] |
| Attenuation | 0.19 $dB/km$ |
| Dispersion | 16.8 $ps/(nm \cdot km)$ |
| PMD | 0.1 $ps/\sqrt{km}$ |
| Nonlinear Index $n_2$ | $2.25 \cdot 10^{-8} \mu m^2/W$ |
| Core Area | 84.95 $\mu m^2$ |
| **ROADM** | |
| Active | when replacing INT |
| Filter Shape | rectangular |
| Bandwidth | channel spacing |
| **Amplifier (Loop)** | |
| Gain | fiber loss |
| Noise | none |
| **Local Oscillator** | |
| Center Frequency | 193.4 $THz$ |
| Linewidth | none |
| **Coherent Frontend** | |
| Pol Splitter | perfect |
| Other Losses | none |
| Hybrid Imbalances | none |
| PD Type | PIN |
| PD Responsivity | 1 |
| PD Dark Current | 0 |
| PD Thermal Noise | none |
| PD Shot Noise | none |

*Figure 4.3: Transmission setup*

*Table 4.3*: *Simulation parameter of optical components. Values in [ ] are only used if explicitly stated in the text.*

## 5 Quality Measures

There is a wide range of measures available concerning optical transmission quality. In this chapter the quantities used to display the results in **Chapter 6** and the reason for their selection is discussed.

### 5.1 Nonlinear Noise Power

The main goal of this investigation is a specific characterization of the influence of NLI, and XMCI in particular, on the deviation of the received from the sent signal. Usually, the *error vector magnitude* (EVM) is used for this kind of measurements. EVM is the root-mean-square of the Euclidian distance between the received sample and the ideal constellation point (length of the error vector), normalized to the mean (sometimes maximum) amplitude of the ideal constellation. **Figure 5.1** shows example error vectors of received symbols (red) and the ideal QPSK symbols (blue). The value the EVM is normalized to in this case is the length of the *symbol vector* (SV, blue).

EVM has two disadvantages: Firstly, it lacks information about spatial distribution of the received symbols in the I,Q-plane. It would require splitting the vector into different directional contributions, which would then be a new measure and not EVM. The second disadvantage is the comparability with the results from the models in **Chapter 3**. GN- and EGN-models yield the PSD of the XMCI induced distortions, so a measure is needed that can easily be calculated by either frequency domain models as well as the time domain deviation from ideal constellation points and can also be split up into directional contributions.

The noise power of the NLIN is such a measure. Starting from the PSD of the distortion, it can be calculated by integrating over the frequencies of the CUT (**Figure 5.2**):

$$P_{XMCI} = \int_{f \in CUT} |G_{XMCI}(f)|^2 \, df \tag{5.1}$$

In the constellation diagrams the same quality can be identified by the variance of the deviation of the received samples ($y_i$) from their ideal constellation points ($x_i$) (**Figure 5.3**):

$$P_{NLI} = S_R \cdot var[(y_i - x_i)] \tag{5.2}$$

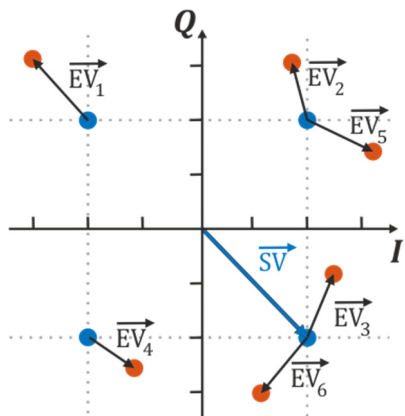

**Figure 5.1:** *Error vectors of the received symbols and ideal symbols. The EVM would be then $\sqrt{<|EV_i|>}/|SV|$*

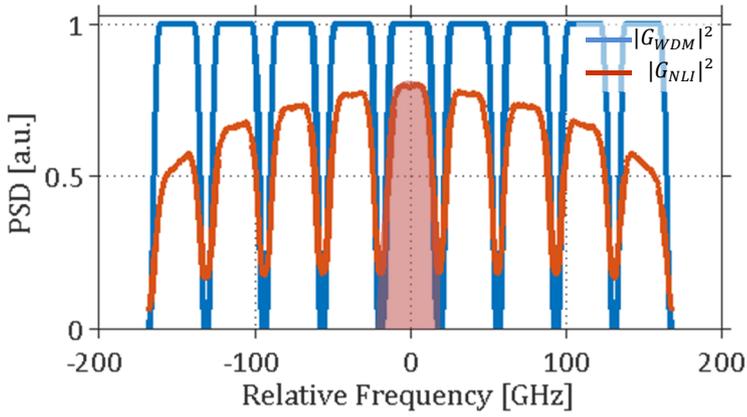
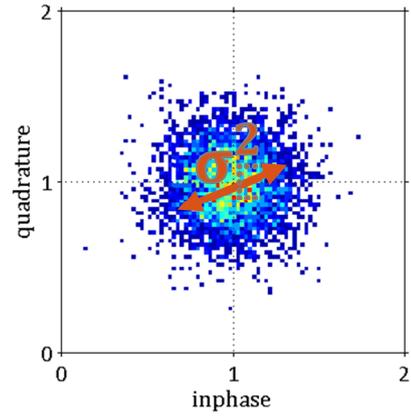

*Figure 5.2: Noise power from PSD. From the PSD of the signal(blue) the NLIN PSD is predicted (red, here: GN-Model). Integrating the latter over f ∈ CUT will yield the NLIN power (red area).*

*Figure 5.3: Noise power from constellation. Sample density is depicted by color-grading. The noise power is retrieved from the variances of the received symbols, with respect to the ideal symbol, multiplied by $S_R$.*

In the simulations it is taken care, that the distortions are caused only by XMCI (see **Chapter 4**), thus when applying Eq.(5.2) on the simulation results $P_{NLI}$ is actually also $P_{XMCI}$.

As mentioned in **Chapter 4**, all simulations were conducted with 5 different PMD realizations of the link, with identical physical parameters, but all random processes initialized differently. It can be seen in *Figure 5.4* that the deviations are negligible, as the individual curves from the realizations do not deviate much from their average – especially in terms of total noise power (left picture). The curves of the noise power and circular noise ratio in **Chapter 6** are still averages over these realizations, though. Another averaging is done over X and Y polarization.

The matter of the distribution of the distortion in the I,Q-plane is handled in **Chapter 5.2**.

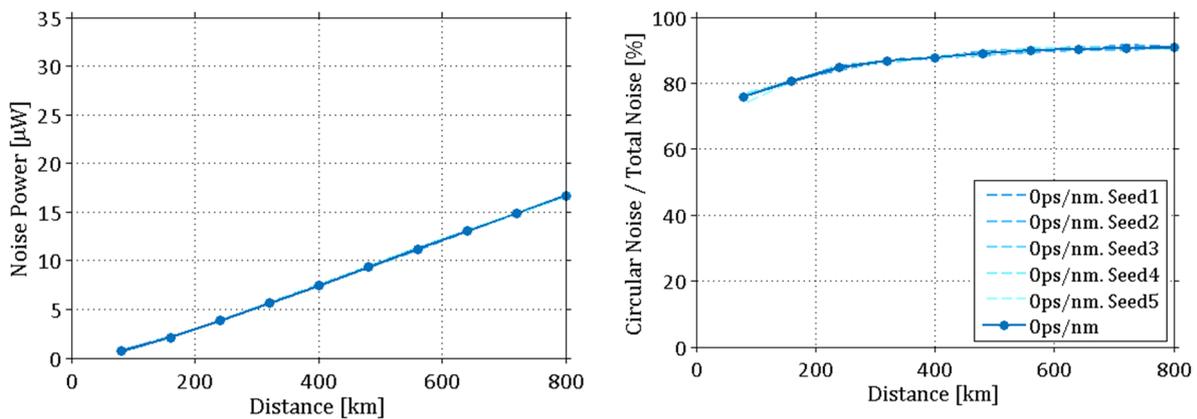

*Figure 5.4: Example of averaged realizations. Simulation results of co-propagating QPSK-INT without pre-dispersion as example. The continuous lines are the averages of the dashed lines, which represent the results for different realizations of the link (already averaged over X- and Y-Polarization). The mean deviation is always less then 1% (mean over distance 0.47%) in case of total noise power and less then 2% (mean over distance 0.78%) in case of circular noise power.*
*The noise ratio is discussed in Chapter 5.2.*

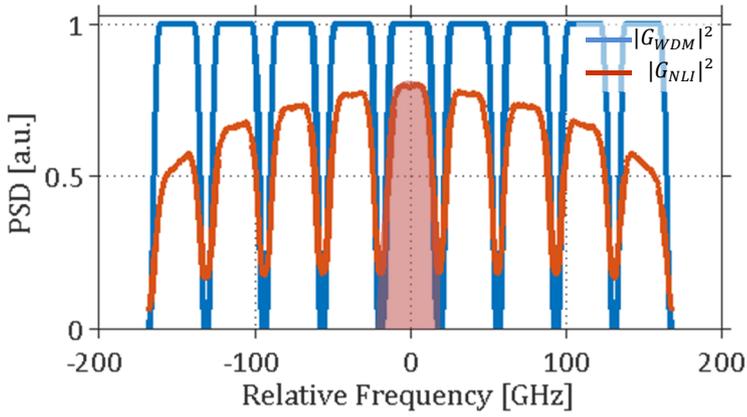
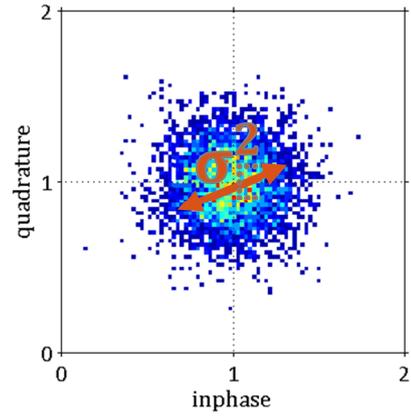

*Figure 5.2: Noise power from PSD. From the PSD of the signal(blue) the NLIN PSD is predicted (red, here: GN-Model). Integrating the latter over f ∈ CUT will yield the NLIN power (red area).*

*Figure 5.3: Noise power from constellation. Sample density is depicted by color-grading. The noise power is retrieved from the variances of the received symbols, with respect to the ideal symbol, multiplied by $S_R$.*

In the simulations it is taken care, that the distortions are caused only by XMCI (see **Chapter 4**), thus when applying Eq.(5.2) on the simulation results $P_{NLI}$ is actually also $P_{XMCI}$.

As mentioned in **Chapter 4**, all simulations were conducted with 5 different PMD realizations of the link, with identical physical parameters, but all random processes initialized differently. It can be seen in *Figure 5.4* that the deviations are negligible, as the individual curves from the realizations do not deviate much from their average – especially in terms of total noise power (left picture). The curves of the noise power and circular noise ratio in **Chapter 6** are still averages over these realizations, though. Another averaging is done over X and Y polarization.

The matter of the distribution of the distortion in the I,Q-plane is handled in **Chapter 5.2**.

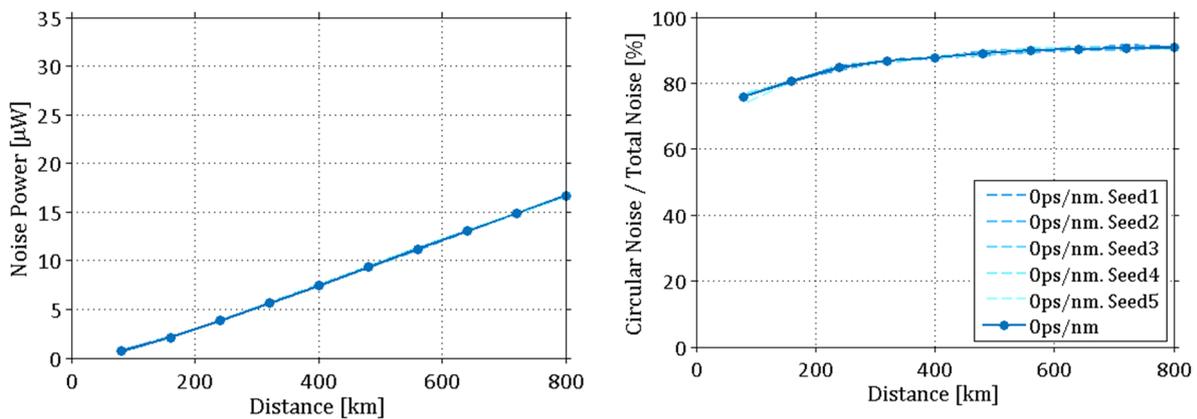

*Figure 5.4: Example of averaged realizations. Simulation results of co-propagating QPSK-INT without pre-dispersion as example. The continuous lines are the averages of the dashed lines, which represent the results for different realizations of the link (already averaged over X- and Y-Polarization). The mean deviation is always less then 1% (mean over distance 0.47%) in case of total noise power and less then 2% (mean over distance 0.78%) in case of circular noise power.*
*The noise ratio is discussed in Chapter 5.2.*

## 5.2 Phase Noise to Circular Noise Ratio

A crucial point of interest is, whether the assumption of gaussianity of the NLIN from the GN-model holds in the investigated cases or the effect of additional phase noise, demonstrated by Dar et al. ([31]), might be influential in the context of flexible optical networks.

Customarily, the separation is made into phase noise and amplitude noise. However, as the Gaussian distributed noise has components in all directions, the division should be done into pure phase noise and the Gaussian part of the noise (see *Figure 5.6*), the latter also called circular noise in the following. The property of multiple symbol durations spanning time correlation of the phase noise (**Chapter 5.3**), was used to separate pure phase noise from the part of the noise that is Gaussian distributed.

Like in the section before, the transmitted symbols are denoted by $x_j$ and the received symbols by $y_j$. The noise contributions $n_j$ represent the Gaussian part of the NLIN. The process is still applicable, when $n_j$ includes additional statistically independent noise contributions of zero mean, for example ASE-noise, which will be obvious from the following steps. The phase difference of the phase noise contribution is denoted by $\Delta\Theta_j$. $y_j$ can be written as

$$y_j = x_j e^{-i\Delta\Theta_j} + n_j \tag{5.3}$$

All constellation points can be brought to a comparable frame by rotating and scaling all symbols according to their ideal phase and amplitude (*Figure 5.5*).

$$\frac{x_j^* y_j}{|x_j|^2} = e^{-i\Delta\Theta_j} + \frac{x_j^* n_j}{|x_j|^2} \tag{5.4}$$

Because multiplication by $x_j^*/|x_j|^2$ is just rotation and scaling, the second term will still have zero-mean property. The first term on the right hand side is the phase noise term and will change only slowly in comparison to the second one, according to theory (**Chapter 5.3**). Hence, averaging over N symbols removes the second term, if N is small enough with respect to the correlation time of the phase noise, but large enough to ensure sufficient statistics.

$$\left\langle \frac{x_n^* y_n}{|x_n|^2} \right\rangle_{n=j-\frac{N}{2},\dots,j+\frac{N}{2}} \cong e^{-i\Delta\Theta_j} \tag{5.5}$$

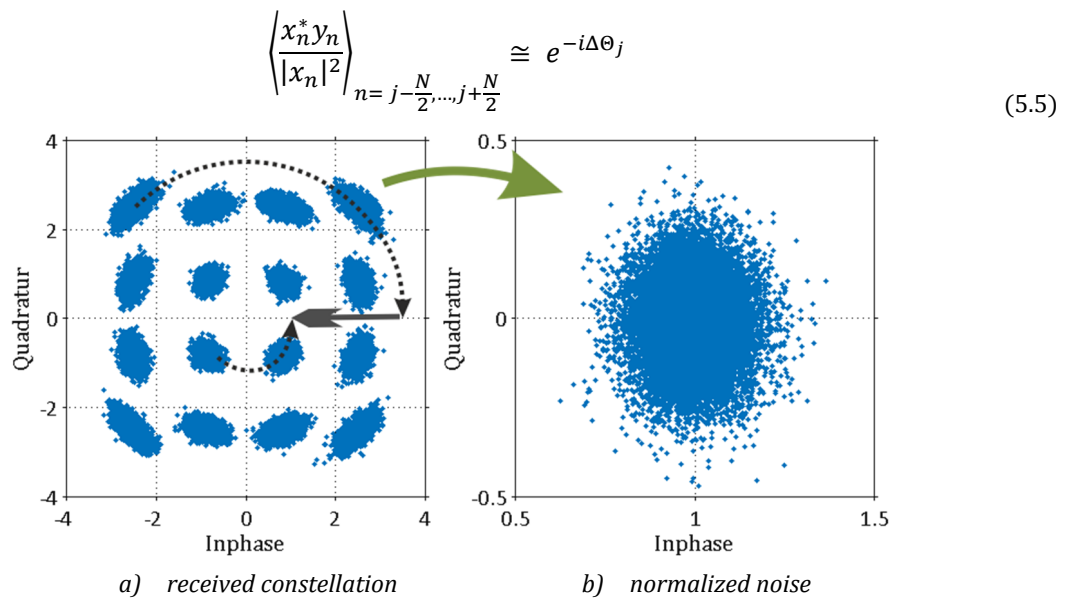

a)   received constellation     b)   normalized noise

***Figure 5.5:*** *Noise normalization. The constellation points are rotated to the real axis and then normalized to unity.*

After using Eq.(5.5) to remove the phase noise and $x_j$ to remove modulation from $y_j$ the circular noise is obtained:

$$n_j^N = y_j - x_j \cdot \left\langle \frac{x_n^* y_n}{|x_n|^2} \right\rangle^*_{n=j-\frac{N}{2},\ldots,j+\frac{N}{2}} \quad (5.6)$$

$$n_j^{circular} \cong n_j^{N_{opt}} \quad (5.7)$$

Choosing N too small will cause the algorithm to remove the phase noise completely, but also phase components originating from the Gaussian noise (**Figure 5.7** red). Choosing N too big on the other hand will eliminate also the phase noise and it will therefore not extracted (**Figure 5.7** blue). Therefore, the variance in quadrature direction ($\sigma_Q^2(N) = var[\Im(n_j^N)]$) is a monotonically increasing function of N. At the same time the inphase component of $n_j^{circular}$ will merely change its variance ($\sigma_I^2(N) = var[\Re(n_j^N)]$).

The monitor signal

$$M(N) = \frac{\sigma_Q^2(N)}{\sigma_I^2(N)} - 1 \quad (5.8)$$

is therefore also monotonically increasing as well. Taking into account that a circular noise distribution will exhibit equal variances in all directions, hence $\sigma_Q^2(N_{opt}) = \sigma_I^2(N_{opt})$, $N_{opt}$ is found when the monitor signal equals zero.

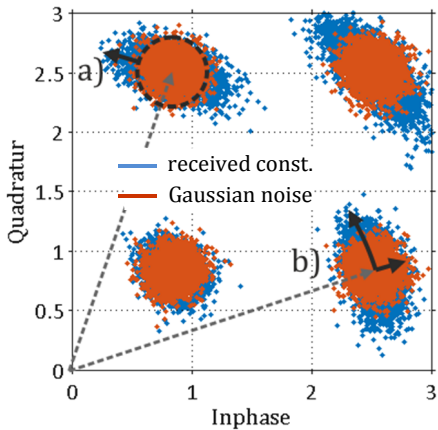

***Figure 5.6:*** *a) Separation into Gaussian- and phase-noise contributions (used to extract Gaussian noise, red). b) Separation into parallel amplitude noise and perpendicular phase-noise.*

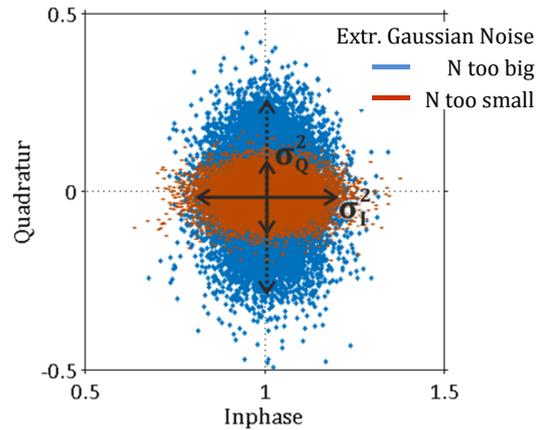

***Figure 5.7:*** *Effect of choosing N either too small (phase of Gaussian contribution is also removed, red) or too big (phase noise is also eliminated and not removed, blue). The optimum N is found when $\sigma_Q^2 = \sigma_I^2$.*

The algorithm implemented takes advantage of the monotonic behavior of $M(N)$ by starting at N=2 and increasing N as long as $M(N) \leq 0$. As soon as $M(N) \geq 0$, $N_{opt}$ is found and the search can be interrupted. In case of dual polarization the algorithm stops when both $M_X(N)$ and $M_Y(N)$ are larger than zero. It will then choose $N_{opt}$ as the point where $|M_X(N) + M_y(N)|$ is smallest. In **Figure 5.8** an example of $M(N)$ for two polarizations is given, stopping when both curves crossed y=0 and finding $N_{opt}$ at 22 taps.

If all noise contributions are circular $M(N)$ will grow asymptotically toward zero and $N_{opt}$ will be the total number of symbols of the simulated signal. To save computation time one can define a maximum number of averaging samples $N_{max}$ or a different stop criterion ($M(N) \geq -eps$ for example).

Having separated phase noise (Eq.(5.5)) and circular noise (Eq.(5.7)) their noise power can be calculated. For $n_j^{circular}$ this is be done analog to (5.2):

$$P_{circular} = S_R \cdot var[n_j^{circular}] \tag{5.9}$$

To arrive at a comparable measure for phase noise, the displacement is assumed to be small so that it can be approximated by $\tan(\Delta\Theta_j) \cong \Delta\Theta_j$. The average phase noise power is then approximated by

$$P_{phase} = S_R \cdot var[\Delta\Theta_j] \cdot E\left[|x_j|^2\right] \tag{5.10}$$

In **Chapter 6** the ratio $P_{circular}/P_{NLI}$ in percent will be plotted, $P_{NLI}$ being calculated according to Eq. (5.2). As $P_{phase} + P_{circular} = P_{NLI}$, $P_{phase}/P_{NLI}$ is then represented by the difference of the shown value to 100%. This is chosen over $P_{phase}/P_{circular}$ to provide equal scaling for the contributions of $P_{circular}$ and $P_{phase}$ without changing to logarithmic units. The value $P_{circular}/P_{NLI}$ will be referred to as *circular noise ratio* (CNR).

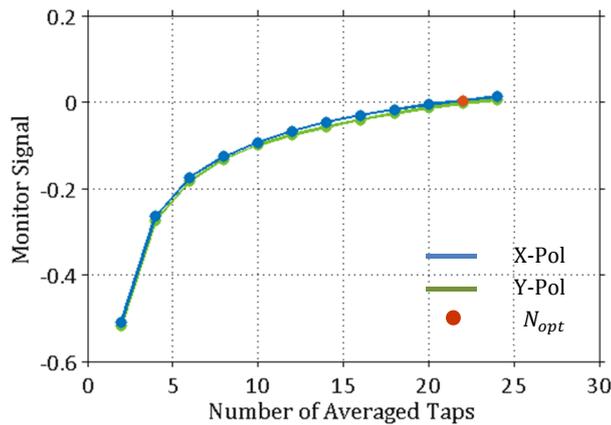

**Figure 5.8:** *Example of the Monitor Signal as described in Eq.* (5.8) *At N=24the monitor M(24) > 0 for both polarizations. $N_{opt}$ is then chosen to be 22, because the mean deviation from zero in both polarization is smallest.*

## 5.3 Time correlation

It was shown in **Chapter 3.2** that not all contributions to the NLIN are randomly Gaussian distributed as assumed in the GN-model, but that there are in fact pulse collisions which will yield only perturbations perpendicular to the symbol vector, namely phase noise. Dar et al. ([31],[37]) point out that the phase noise contributions $\Delta\Theta_j$ (**Chapter 5.2**) are correlated in time, meaning the discrete *auto-correlation function* (ACF), Eq.(5.11), will not vanish at the delay of one symbol, but is actually correlated for a number of consecutive symbols, depending on the system parameters.

$$ACF[\Delta\Theta_j](x) = \sum_j \Delta\Theta_j \cdot \Delta\Theta_{j+x}, \qquad x \in \mathbb{Z}$$
(5.11)

It is clear, that at $x = 0$ the ACF has its maximum. For comparability, this value is normalized to unity in the figures.

The property of a non-negligible coherence length can be used to mitigate NLIN [38] and the same process can be utilized to separate the circular Gaussian-noise contributions from phase-noise contributions as shown in **Chapter 5.2**.

*Figure 5.9* shows a plot of the ACF of the ΔΘj. Two qualities of the plots are visible at once: Firstly, the function is symmetric with respect to the y-axis and secondly there are only slight differences in X and Y polarization. Therefore, just the components in X-polarization with $\Delta > 0$ are presented in **Chapter 6**. Furthermore, as only tendencies are investigated in this context and there are only slight differences between the ACF of the realizations, only one realization is shown.

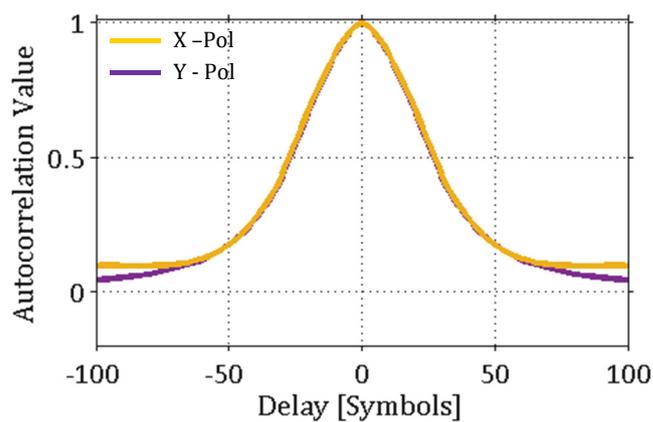

*Figure 5.9: Example visualization of the phase noise ACF.*

# 6 Results and Discussion

In this chapter the results of the simulations are presented, discussed and compared to the analytical models from **Chapter 3**.

At first the influence of accumulated CD is investigated, either being accumulated in the course of transmission or manually imposed on the INT before propagation. Further, the study includes restricting the accumulation of CD on the INT to one span, by replacing the INT after each span with their state at the beginning of transmission. The opposite case, continuing co-propagation of the channels, is therefore also titled 'unrestricted'. The additional accumulated CD on the neighboring channels is also referred to by 'pre-dispersion of the INT', as the INT have acquired this property before being multiplexed with the CUT. There is no case presented, where the CUT is pre-dispersed, hence 'pre-dispersion' and 'additional accumulated CD' always refer to the INT, even if not explicitly mentioned. While examining the influence of accumulated dispersion, also the modulation format influence is investigated. Again, only the modulation format of the INT is changed, while the modulation format of the CUT is always 16QAM.

In the second part of this chapter the influence of span-length on the NLIN is under test.

The third part is concerned with the spacing of the channels.

## 6.1 Accumulated Chromatic Dispersion and Modulation Format

It was mentioned in **Chapter 3** that CD influences different aspects of the noise built-up: Firstly, it leads to broader pulses and thus greater overlap. Secondly, CD also changes the amplitude statistics of the signal. It can be distinguished between the build-up of CD on the INT before they are multiplexed with the CUT and the build-up during co-propagation with the CUT.

To provide an overview of the influences, the following combinations are under test:

- A) none of the channels have acquired CD beforehand and propagate together along the link, so their accumulated CD is always identical. This scenario corresponds to a point-to-point network setup.
- B) the CUT is multiplexed with INT already dispersed by an amount of accumulated CD corresponding to approximately 800km of propagation (13000ps/nm). It this case the difference in accumulated CD is still constant, but with the INT being much more dispersed. In an equivalent network setup the CUT would be generated at a ROADM site and would be received at the next node.
- C) same start conditions as in A), yet, instead of co-propagating over the complete link, the INT are reset after each span to their original state at the beginning of the link. As a result the CUT acquires more CD in each span and the accumulated CD difference between CUT and INT increases during propagation. This corresponds to a flexible network in which all neighboring channels would be newly generated at the node sides.
- D) same start conditions as in B), but analog to C) the INT are reset after each span. In this case however, the INT possess always more accumulated CD than the CUT, but the difference is decreasing. This case aims to represent a more realistic flexgrid scenario, where the neighboring channels have propagated for some time before being joined with the CUT, but will also be replaced at the switching points.

The picture of **Chapter 3.2** additionally suggests a strong dependence on modulation format in terms of noise power as well as CNR. The formats QPSK and 16QAM (*Figure 2.1*) were chosen for investigation, primarily due to their increasing number of amplitude levels, mentioned as a crucial property influencing NLI, but also because of their widespread use in current transmission links. Furthermore interferer with Gaussian distributed constellations were investigated, pushing the signal statistics to the limit of the GN-Model. These three modulation formats ensure completeness of the investigation with regards to modulation format dependence, as the models suggest that the terms depending on modulation will vanish for formats with constant envelope, such as QPSK, and reach maximum influence for Gaussian interferer.

The scenarios are set up as described in **Chapter 4** and compared with the predictions of GN- and EGN- model quantitatively and the PC-picture qualitatively.

—

First, QPSK INT are examined. According to the PC-picture no contributions from two- and type B three-pulse collision are present for this modulation format.

The results of case A) are depicted by the continuous blue curves in ***Figure 6.1*** and ***Figure 6.2***, showing increasing NLIN with transmission distance. Starting with no accumulated CD most of the pulse collisions are complete during the first span. Lacking contribution from two-pulse collisions, according to the PC-picture, only contributions from complete type A three-pulse collisions as well as four-pulse collisions are present, the latter being dominant. This coincides with the CNR of about 78%. As described in the PC-picture the strength of complete collisions of this type is small, hence the small NLIN power. Becoming incomplete during propagation along the link, due to CD accumulation on INT and CUT, especially the noise power of four-pulse collisions should increase. This can be observed by the increasing slope of the curve and the bigger circular noise contribution. The effect seems to saturate, as the power curve becomes linear and the slope of the CNR decreases. The EGN-model, while giving no information about the type of noise, predicts the trend of the power curve very well. In contrast, the GN-model overestimates the noise power from the beginning and overestimates also the accumulation in the following spans, resulting in a steeper curve and therefore increasing deviation with transmission length.

When imposing accumulated CD on the INT, the PC-picture predicts collisions of three pulses, involving two pulses from the INT, right from the start. These kinds of collisions should result in phase noise, which can be observed in ***Figure 6.2***. Here the blue dashed curve represents the case B) from above. In this case, in contrast to A) (blue continuous curve), the most power was added in the first span, but from the subsequent progression of this curve it is clear that the development of both curves are governed by the same NLI contributions: The blue power curves start to run parallel and the difference in their CNR decreases. The only difference of B) compared to A) seems to be the increased noise power in the first span due to the additional phase noise. Hence, when transmitting a channel without dispersion, this channel will be distorted by phase noise, even more so, if the neighboring channels have already accumulated dispersion in the course of their transmission. The further course of the curve indicates that in case of co-propagating channels the influence of pre-dispersion decreases with transmission distance and the NLI is then mainly caused by four-pulse collisions.

The GN-model yields better results for the first span then for the case A), yet still overestimates the NLI formation of the following spans by far.

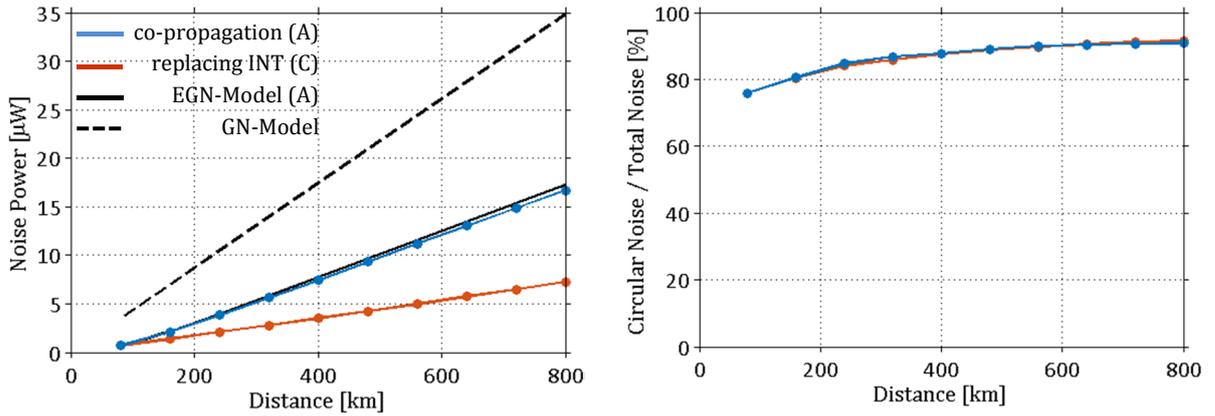

*Figure 6.1:* QPSK INT without acc. Disp. on INT. XMCI total noise power and circular noise percentage over transmission distance, simulation and analytical model results.

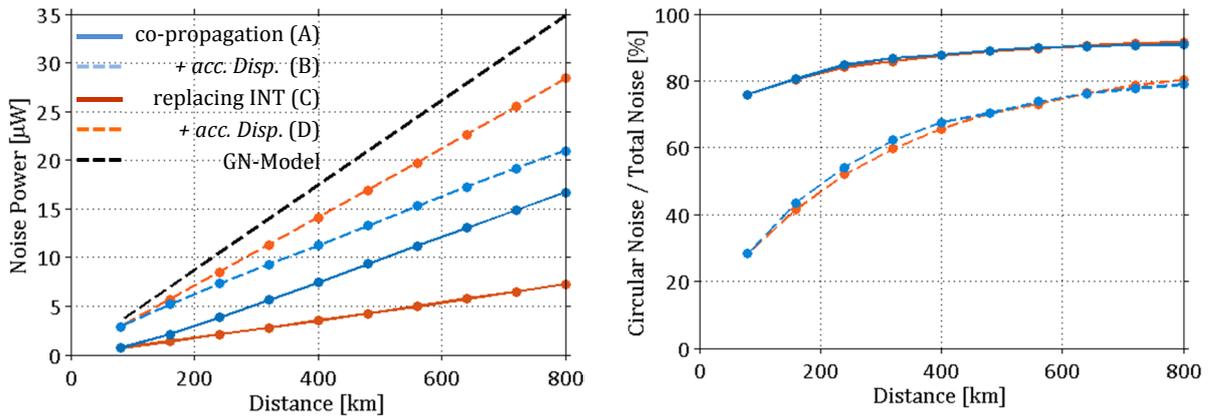

*Figure 6.2:* QPSK INT with and without acc. Disp. on INT. XMCI total noise power and circular noise percentage over transmission distance. The amount of acc. dispersion on the INT is $13000\frac{ps}{nm}$.

Upon resetting the accumulated CD on the INT to 0ps/nm (case C), continuous red curves in *Figure 6.1*) the first thing to note is the constant slope of the curve: The amount of NLIN added is equal in each span (see also *Figure 6.5*). Nonetheless, the trend of the CNR is identical to case A).

In case D) the accumulated CD on the INT is reset after each span to 13000ps/nm. This increases NLI (red dashed curve in *Figure 6.2*) compared to the other cases, the CNR behavior however is almost identical to the curve from B). Similar to case C), the amount of NLIN added in each span does not change, leading to linear curve behavior and the most NLIN generated of the results so far.

To conclude the investigation of QPSK interferer: In case of co-propagating channels the amount of NLIN added after each span is independent of the accumulated CD on the INT for large transmission distances. In case of pre-dispersed INT, additional phase noise is added only during the first spans. Replacing the INT will cause the generation of NLIN of equal power in each span, but the CNR development will be identical to the co-propagating case.

—

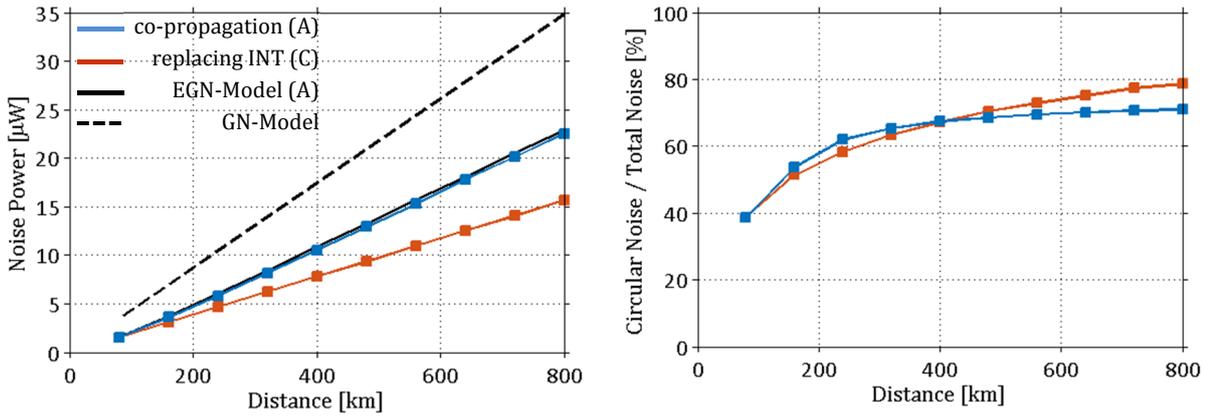

*Figure 6.3:* *16QAM INT without acc. Disp. on INT. XMCI total noise power and circular noise percentage over transmission distance, simulation and analytical model results.*

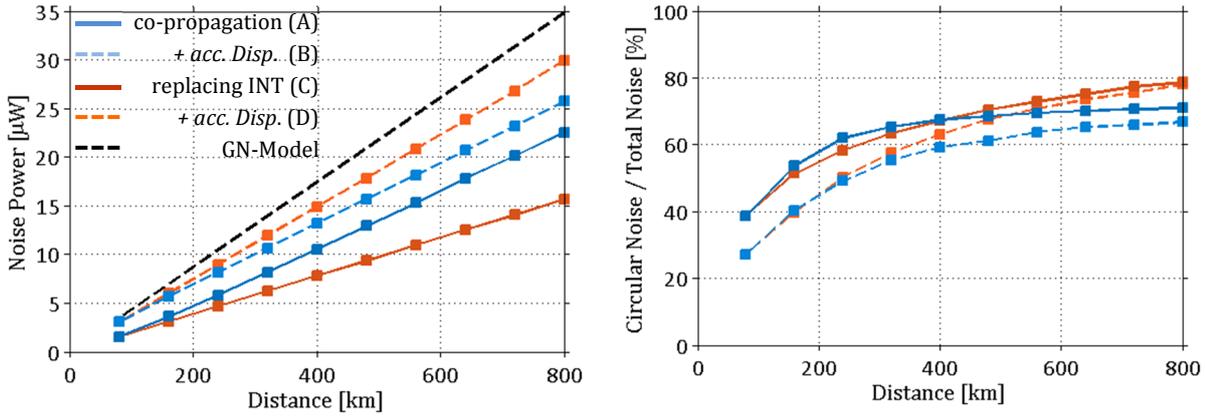

*Figure 6.4:* *16QAM INT with and without acc. Disp. on INT. XMCI total noise power and circular noise percentage over transmission distance. The amount of acc. dispersion on the INT is $13000\frac{ps}{nm}$.*

The same investigation is performed with 16QAM INT. Here, the PC-picture predicts the influence of all four kinds of pulse collisions, yet two- and type B three-pulse collisions with smaller $X_{hkm}$-coefficients compared to Gaussian INT, due to the dependence on the moments.

Two-pulse collisions result in a phase noise contribution, therefore after the first span of propagation in case A) the NLI should be not only bigger than for QPSK INT, but should also have a more pronounced phase noise character. Similar to QPSK INT, after few spans the pulse collisions should be getting more and more incomplete and the phase noise ratio should be lessened by the additional circular noise arising from four-pulse collisions. The final ratio however is supposed to be shifted toward a higher phase noise contribution by the remaining two-pulse collisions. These predictions of the PC-picture coincide with the simulation and are reflected in the progression of the continuous blue curves in ***Figure 6.3*** and ***Figure 6.4***. The gap in noise power to case B) is less pronounced than in the case of QPSK INT.

Adding accumulated CD on the INT, as described in case B), results in more noise power, shown in ***Figure 6.4***, and the phase noise ratio after one span is even more increased. These phenomena can be attributed to the interpretation, that three-pulse collisions of type A are now also present, triggered by widening of the INT pulses. Consequently, the CNR difference to A), decreases as CD accumulates during propagation: the blue curves of the total noise power run parallel, but with an offset caused by their difference during the first span. The CNR approaches the CNR curve of case A) with transmission distance, closing the gap due to the higher percentage of phase noise at the beginning, which is smaller than in case of QPSK INT.

# 6 Results and Discussion

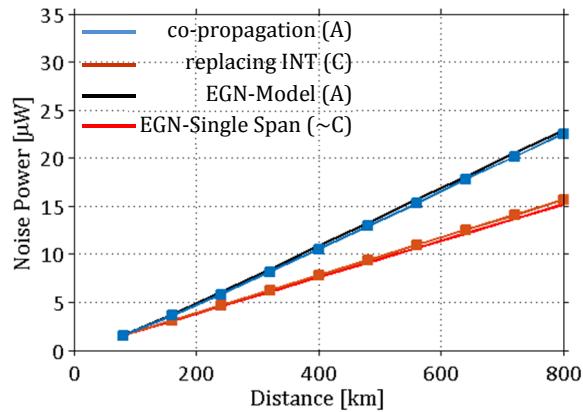

***Figure 6.5:*** *16QAM INT and no acc. Disp. on INT. XMCI total noise power. A modified EGN-Model is shown in light red, simply multiplying the NLI from the first span with the number of spans.*

In case of resetting the accumulated CD on the INT (C), continuous red curve in ***Figure 6.4***) type A three-pulse collisions and four-pulse collisions should be suppressed, because they rely on two overlapping pulses from the INT. This expectation is met by the fact that the NLIN added in each span is less than in A). Meanwhile, the CNR indicates that mainly phase noise is suppressed. The CNR curves show a convergence with the curve of D) (red dashed curves in ***Figure 6.4***), which was not visible in the case of QPSK INT, due to the large gap between the CNR curves. The slope of the power curves remains the same after each span, for both cases.

When comparing the results with GN- and EGN- model the same conclusions as in the case with QPSK INT can be drawn. Additionally, it was investigated whether the EGN-model can be quickly adapted to reproduce the effect of replacing the INT, which was simply done by using the results of the first span for each following span. The well-fitting result can be seen in ***Figure 6.5***.

—

The examination of modulation formats is completed with the study of Gaussian INT (***Figure 6.6*** and ***Figure 6.7***). The trends mentioned in comparing QPSK INT with 16QAM INT are at their limits: The curves in power and CNR are at the same value after the first span. Also, resetting the dispersion after each span (red curves) does not make a difference for Gaussian INT in terms of noise power, the slope of the power curves is almost identical. In terms of CNR, unlike QPSK INT but similar to 16QAM INT, the red curves of reset INT now follow each other, instead of their blue counterparts and also to a higher ratio of circular noise.

The PC-picture gives an explanation for the increased phase noise ratio in comparison with QPSK INT or 16QAM INT, as the modulation dependent factor reaches its largest value, leading to maximum contribution of two-pulse collisions and therefore increased phase noise.

Summarized: In case of Gaussian INT there is only a change of outcome if the accumulated CD on the INT is either held constant or if it grows with propagation distance. The quantity of accumulated CD is of no importance. Moreover, this change is only visible in the ratio of phase and circular noise, not in the noise power.

The GN- and EGN- model, only differing in the modulation format of the CUT, result in almost identical curves, suggesting very small correction terms. Their results are in agreement with the simulation.

—

***Figure 6.8*** shows the phase noise autocorrelation functions for 16QAM INT at different stages of the link for the different setups A) to D). It can be seen, that the correlation of the phase noise changes with transmission length, pre-dispersion and resetting the INT. While correlation is build up during propagation – visible from the higher blue curve taken at 800km transmission length to the yellow curve at only 80km -, replacing the INT results in a reduction of this correlation, obvious from the red curve. This is due to the random phase rotations and delays performed on the replaced INT: While in co propagating channels the phase relations between frequency components vary only due to phase noise, which is small, the phase relations between CUT and INT frequency components are completely altered by these random variations, leading to uncorrelated phase noise. Pre-Dispersion on the INT (b)) also reduces the autocorrelation, but by a much smaller amount compared to the reset of the INT.

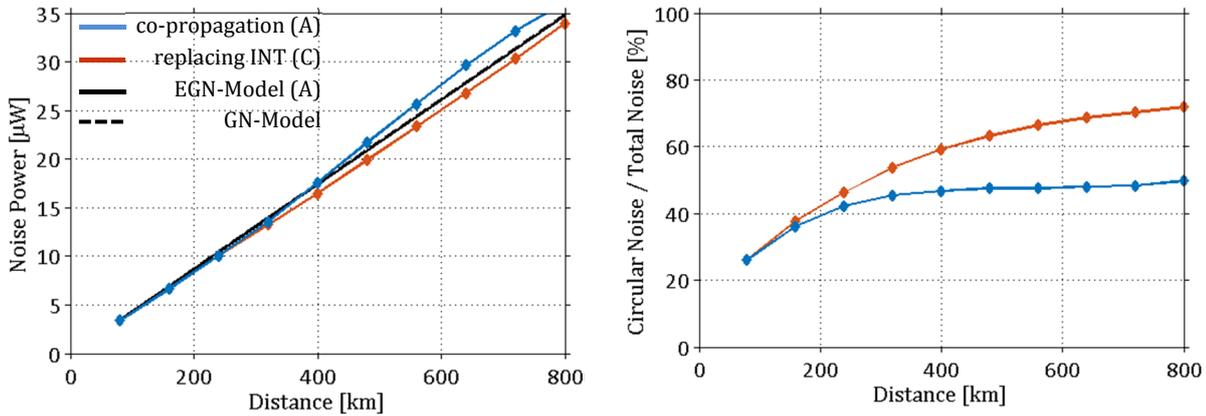

*Figure 6.6: Gaussian INT without acc. Disp. on INT. XMCI total noise power and circular noise percentage over transmission distance, simulation and analytical model results.*

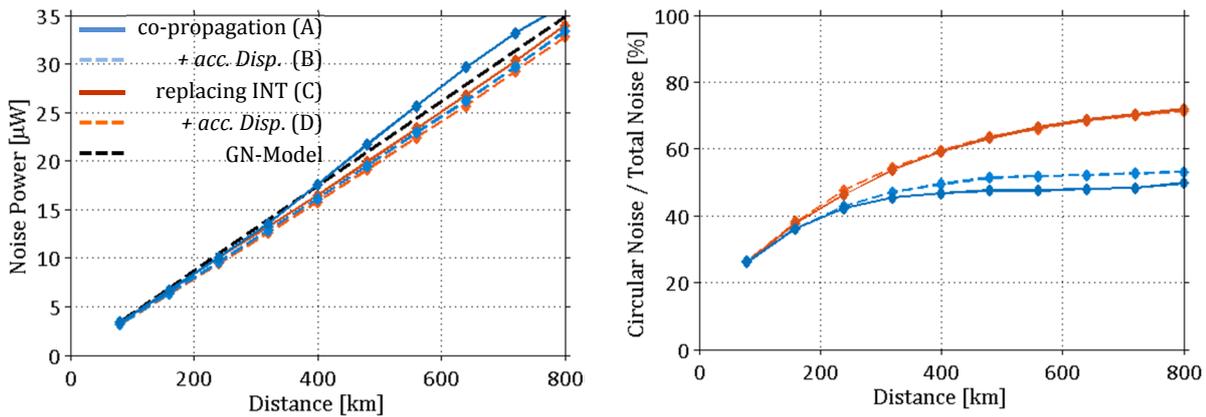

*Figure 6.7: Gaussian INT with and without acc. Disp. on INT. XMCI total noise power and circular noise percentage over transmission distance.. The amount of acc. dispersion on the INT is $13000\frac{ps}{nm}$.*

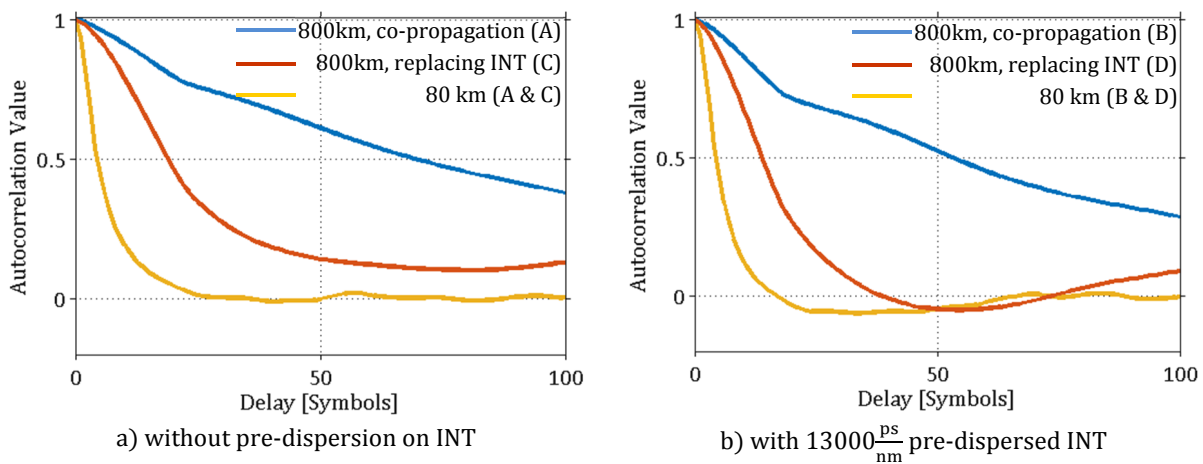

a) without pre-dispersion on INT    b) with $13000\frac{ps}{nm}$ pre-dispersed INT

*Figure 6.8: Phase Noise ACF of simulations with 16QAM INT after 1 Span (80km) and 10 Spans (800km).*

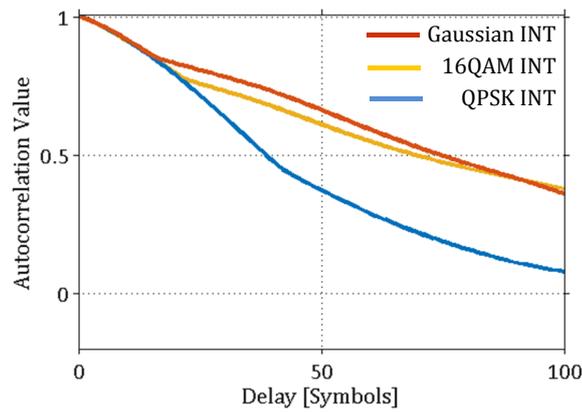

*Figure 6.9: Phase Noise ACF after 800km co-propagation, no pre-dispersion. Results shown for the different modulation formats on the INT.*

Comparing different modulation formats (**Figure 6.9**), it can be seen, that in case of Gaussian INT (red) the correlation declines slowest, while QPSK INT (blue) yields the fastest drop in correlation. The curves differ not so much in their slope, but mainly in the position of the kink and therefore in the height of tails.

Concluding: the possibility to mitigate the NLI phase noise by averaging is diminished by exchanging the INT; in case of QPSK INT even more so, when the INT have already acquired acc. CD.

—

This study supports the claim of [9] stating that adding accumulated CD to the INT will not make the signal more Gaussian, which was suggested in [26], as the blue dashed curves in **Figure 6.2**, **Figure 6.4** and **Figure 6.7**, representing co-propagation and pre-dispersion, differ essentially from each other and are in fact rather similar to their continuous counterparts without pre-dispersion. Yet, when adding accumulated CD and keeping it limited by resetting the INT, the results actually seem to be of more Gaussian nature, hence the red dashed curves to not differ much from each other. Consequently, the GN-model seems to be a fast and reasonable estimator for the power of NLI in flexible optical networks. The EGN-model on the other hand needs to be consulted, when more accurate assessments are needed or in case of networks of co-propagating channels.

The authors of [9] also claim, that instead of the signal becoming Gaussian the pulse collisions become incomplete. This on the other hand does seem to explain the behavior of the NLIN caused by co-propagating INT, but fails to explain the increase in noise power in case D) for QPSK INT and 16QAM INT. This also fails to explain the convergence of CNR of co-propagating channels with and without pre-dispersion in case of Gaussian INT. Also, even with pre-dispersion, the curves of co-propagating channels in case of QPSK and 16QAM INT differ substantially from the GN-model.

## 6.2 Span-Length

A further parameter under investigation is span length. For reasons of brevity only the case of 16QAM INT is examined, as it is a mixture of tendencies visible for QPSK INT and Gaussian INT as seen in **Chapter 6.1**.

The nonlinear regime of an attenuated span is classified by the effective length $L_{eff}$, which corresponds to the length of a lossless fiber with equal power-length-product (see Eq. (2.18)). If the span length is reduced from 80km to 40km, keeping all other parameters identical, $L_{eff}^{80km}$ goes down from 22.2km to $L_{eff}^{40km}$ = 18.9km. Therefore, by cutting the transmission distance in half the NLIN is expected to decrease by only 15% per span. Comparison of the difference between the continuous lines in *Figure 6.10* (blue: 80km spans, green: 40km spans) shows convergence to this value after a few spans. Additional accumulated CD on the INT (dashed lines) reduces the difference; in fact, the noise build-up during the first spans is nearly identical in the pre-dispersed case. Independent of accumulated Dispersion, the CNR is reduced in case of 40km-spans.

It is clear, that when comparing not span by span but by travelled distance the total noise would almost double (minus ~15%) coming from 80km/span to 40km/span, which is easy to see by comparing the curves at 800km, corresponding to 10 spans of 80km-spans or 20 spans of 40km-spans.

The GN-model still overestimates the NLIN generation, while the EGN-model yields results much closer to the simulated values. This is similar to the case of 80km-spans, but here the model slightly underestimates the noise power with increasing distance.

A different picture emerges, when keeping the accumulated dispersion on the INT constant. The results of this simulation are shown in *Figure 6.11*. When the accumulated CD is identical at the beginning of each span, the noise power contribution of the span stays the same, as it was the case for 80km-spans. Therefore, as the power contribution of the first span is almost the same for 40km- and 80km-spans, their curves in the case of resetting the INT stay close together.

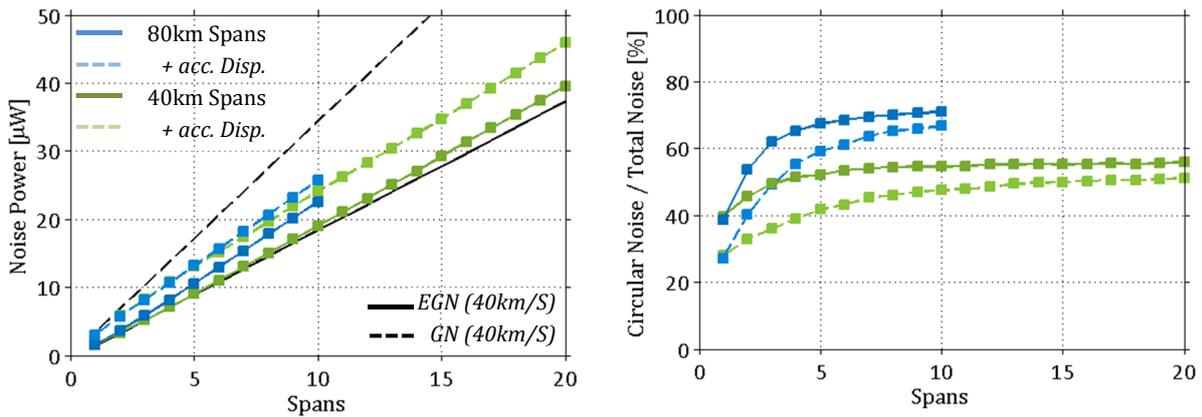

*Figure 6.10:* Co-Propagating 16QAM INT with and without acc. Disp. on INT for different span length. XMCI total noise power and circular noise percentage over transmission distance. The amount of acc. CD on the INT is $13000\frac{ps}{nm}$.

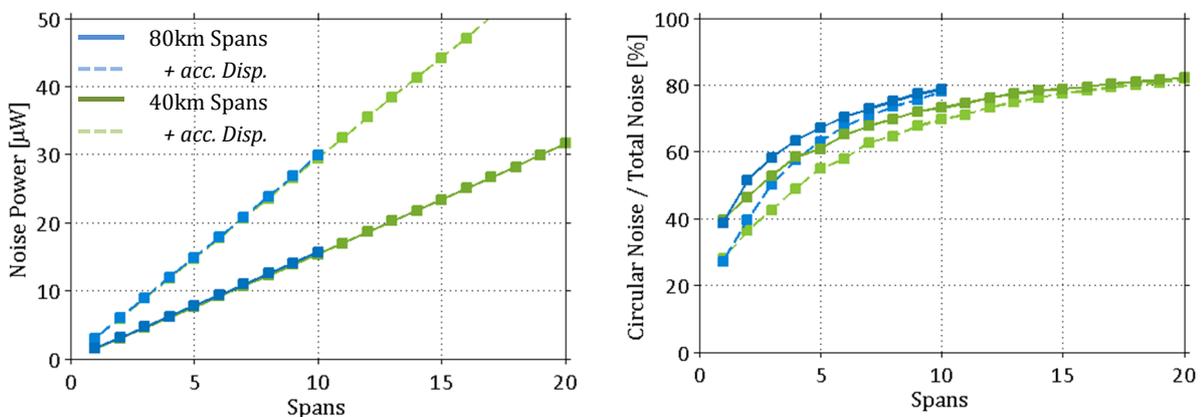

*Figure 6.11:* Replaced 16QAM INT with and without acc. Disp. on INT for different span length. XMCI total noise power and circular noise percentage over transmission distance.. The amount of acc. CD on the INT is $13000\frac{ps}{nm}$.

# 6 Results and Discussion

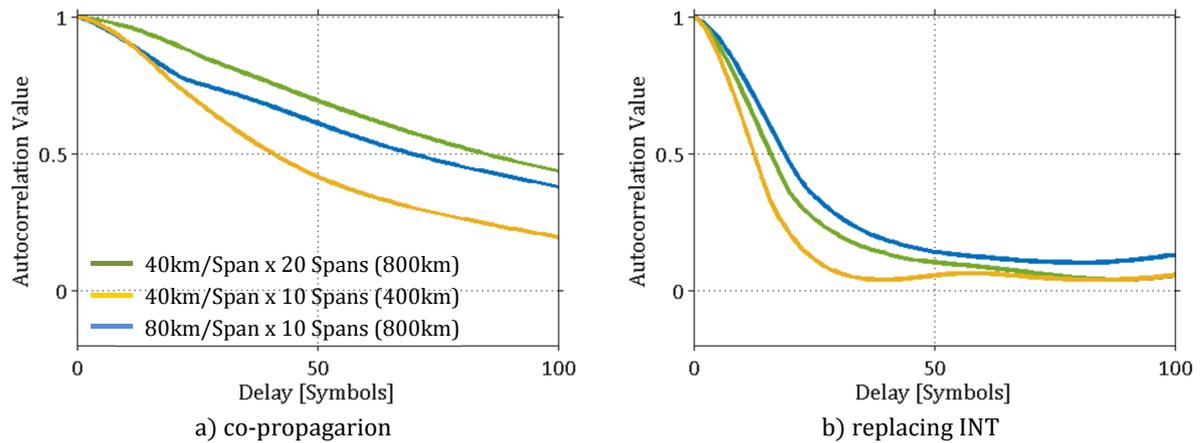

a) co-propagarion | b) replacing INT

***Figure 6.12:*** *Phase Noise ACF for co-propagating 16QAM INT, without acc. Disp. on INT for different span lengths after 1, 10 and 20 spans.*

The increase of the CNR on the other hand seems to be related to the increase in accumulated CD on the CUT and is smaller span-wise, yet similar kilometer-wise. The CNR curves converge to almost the same CNR as for 80km. Therefore in this case of replaced INT there is no additional phase noise contribution as there was with co-propagating channels.

A span-wise comparison of the phase noise ACF of co-propagating channels ***Figure 6.12*** a), yields that the correlation of the phase noise decreases with the use of 40km spans (blue vs. yellow), even though the phase noise ratio was increased. Yet, on the other hand, it increases slightly when comparing at 800km (blue vs. green). Both behaviors can be attributed to the difference of $L_{eff}$: The first is a result of the shorter $L_{eff}$ per span, the latter a result of the almost doubled $L_{eff}$ per transmitted km during which the correlation can build up.

In contrast, when replacing the INT (***Figure 6.12*** b)) the correlation per span is reduced, as it was the case with 80km spans, but also the ACF at 800km (green) is now even lower than that of 800km with 80km spans (blue).

This supports the explanation given in **Chapter 6.1** for the loss of correlation due to INT replacement: With 40km-spans the correlation is broken by the random delays and phase rotations double as often as in the case of 80km-spans, destroying the correlation of the phase noise, even though it is reaches a slightly higher percentage value.

Therefore, using shorter spans not only increases NLIN per kilometer due to almost doubled $L_{eff}$, but it also reduces the possibility to mitigate phase noise in case of replaced INT. In co-propagating systems on the other hand one might be able to remove more of the NLIN for two reasons: Firstly the amount of phase noise is larger in comparison with longer spans and secondly the time-correlation of this phase noise is increased.

## 6.3 Channel Spacing

The last parameter under investigation is the spacing between the channels. Looking at it from the GN-model point of view of **Chapter 3.1** the distance between the spectral components is changed, varying the weights $\mu(f_1, f_2, f)$. In a pulse collision context the walk-off $\beta_2 \Omega_w z$ of the pulses from different channels is influenced by their spacing. Both predict less NLI with increased channel spacing. The spectral width of the signal occupying the channels is 33.6GHz, hence the channel spacing of 37.5GHz used so far leaves little room for reduction of the gap between channels. Instead, in agreement with ITU standards (ITU-T Rec. G.694.1) the width of each channel is increased in steps of 12.5GHz and the spacings under investigation are therefore 37.5GHz, 50.0GHz and 62.5GHz.

From ***Figure 6.13*** it is clear, that the prediction of the models is correct: channels placed farther apart cause less NLIN on the CUT. The non-linear decrease was also expected by both, PC- and (E)GN- model. In fact, no closed form dependence on channel spacing can be given by either of the models, yet the PC- picture gives the dependence in the regime of complete collisions on frequency spacing as linear decreasing in case of two-pulse collisions, quadratic decreasing for three-pulse collisions and cubic decreasing in case of four-pulse collisions. During the first spans there is therefore a decrease of circular noise expected with increasing channel spacing. In the regime of incomplete collisions, occurring with increasing transmission distance, all types of pulse collisions are predicted to scale equally with frequency distance. The convergence of the CNR-curves at larger distances is in agreement with the PC-picture. The EGN-model predictions are also in very good agreement with the power curves of the simulations for all of the investigated channel spacings.

Increasing the spacing between channels also increases the phase noise ACF, especially in the tail of the function, which can be seen in ***Figure 6.14***.

A main goal in flexible optical networks is to increase spectral efficiency, which translates to packing the channels as close as possible. The results shown before indicate thus increasing NLIN.

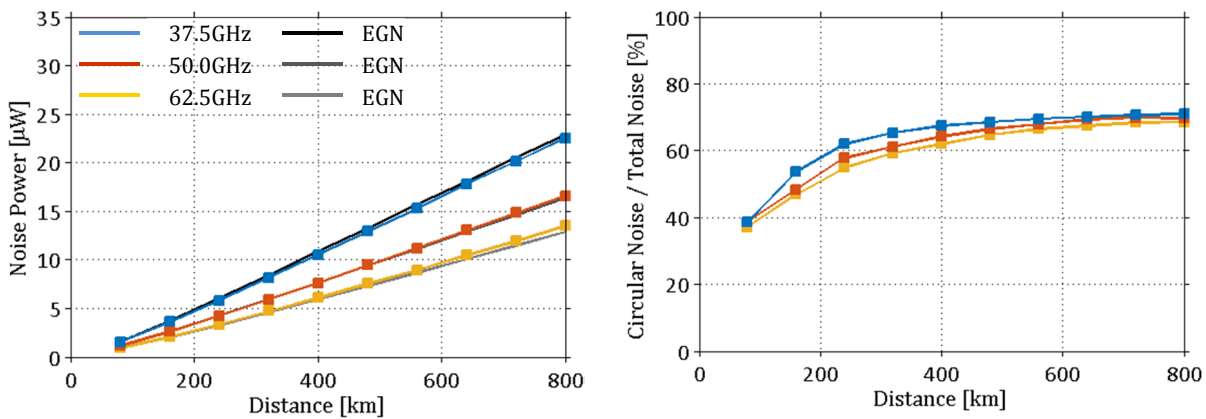

***Figure 6.13:*** *XMCI total noise power and circular noise percentage over transmission distance. Co-Propagating 16QAM INT without acc. Disp. on INT for different channel spacings.*

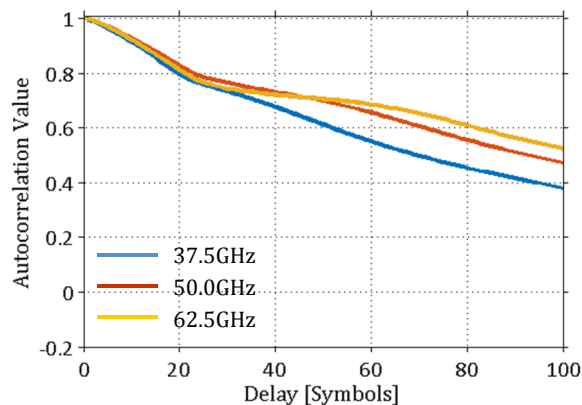

***Figure 6.14:*** *Phase Noise ACF for co-propagating 16QAM INT after 800km, without acc. Disp. on INT for different channel spacings.*

# 7 Conclusion

In the course of this thesis the properties of NLIN and its dependence on system parameter in flexible optical networks were presented and discussed. To do so, first, basic background information about flexible optical networks and signals therein was given. Secondly, analytical models to calculate NLIN, as well as a model to interpret the process of NLIN formation, were introduced. Thirdly, the numerical simulation setup was described and the precautions taken, to extract distortions caused by XMCI only, pointed out. Then, the measures used to define transmission quality were discussed.

Finally, the results of the thesis were presented. It was shown that adding accumulated CD on the INT increases noise power formation during the first span and this additional noise is mainly of phase noise character. When the INT are reset to their original state after every span, the noise generated during each span stayed the same, yet the ratio of phase noise and circular noise changed for multi-amplitude level INT. Additionally, multi-amplitude-level modulation formats on the INT increase NLIN formation, with the maximum for Gaussian distributed levels. The phase noise ratio was also increased in these cases.

Reducing the length per span, while keeping the number of spans identical, reduces NLIN only in case of co-propagating channels. The phase noise ratio on the other hand is increased in both cases.

Spacing the channels farther apart reduces the noise power but has little effect on the ratio of the noise contributions, except for an increase of phase noise ratio is increased during the first spans.

The phase noise correlation time is shortened with additional dispersion and resetting the INT after each span, but increased the longer the non-linear co-propagating length of the link. Larger channel spacing increases phase noise correlation, too. As in flexible optical networks the channels are often replaced at the nodes and are packed closely to increase link capacity, mitigation by noise averaging is constricted.

Insight into the origins of NLI was provided by the PC-picture, which helped in understanding tendencies observed in the results. Where the implemented EGN model could be adapted to the system setup, its results were in agreement with the simulation. The GN- model on the other hand always overestimated the NLIN generation. Deviation from the simulative results was smallest in cases where pre-dispersed INT were replaced after each span. It is the author's opinion, that to quickly assess the transmission quality of flexible optical networks, the GN model is a good choice, yielding fast results and leading to conservative worst-case estimations.

# Acknowledgements

I would like to thank Dr. Martin Schell for supervising this thesis and for the insightful comments and fruitful discussions.

My gratitude goes out to Felix Frey for sharing his fascination for, actually, almost everything in the context of optical signal transmission and his patience in answering the questions I had to almost everything in the context of optical signal transmission.

A heartfelt thank-you to Dr. Carsten Schmidt-Langhorst and Dr. Robert Elschner for their open, extensive and helpful criticism. Additional thanks to the former, for co-supervising this thesis.

I am grateful to Dr. Johannes Fischer for the opportunity to work in the Digital Signal Processing Group of the Photonic Network Department of the Heinrich Hertz Institute.

I would like to thank the other members of the Institute, whom I had the chance to meet, especially to the other master students, whose curious questions were often inspiration for further investigation.

As a final point, I would like to apologize to everybody deprived of my attention during the last few months – it will return to you, immediately after writing this final sentence of this thesis.